\documentclass[useAMS,usenatbib]{mn2e}

\usepackage{makeidx}
\usepackage{rotating}
\usepackage{graphicx}
\usepackage{lscape}

\newcommand{\MS}[0]{M$_{\bullet}$--$\sigma$\ }
\newcommand{\kms}[0]{km~s$^{-1}$}
\newcommand{\mum}[0]{$\mu$m}
\newcommand{\dd}[0]{\textrm{d}}

\newcommand{\apj}{ApJ}
\newcommand{\apjl}{ApJ}
\newcommand{\apjs}{ApJS}
\newcommand{\aj}{AJ}
\newcommand{\mnras}{MNRAS}
\newcommand{\aaps}{A\&AS}
\newcommand{\aap}{A\&A}
\newcommand{\pasp}{PASP}
\newcommand{\procspie}{Proc.~SPIE}

\def\E{{\cal E}}
\def\b#1{{\bf #1}}
\def\d{{\rm d}}
\def\p{\partial}

\def\Lgh{L_{\rm GH}}
\long\def\crap#1{}

\def\psf{{\rm psf}}
\def\Lhist{L_{\rm H}}

\title[]{The central kinematics of NGC 1399 measured with 14pc resolution}
\author[R. C. W. Houghton, J. Magorrian, M. Sarzi, N. Thatte, R. L. Davies, D. Krajnovi\'c]{R. C. W. Houghton$^{1}$, J. Magorrian$^{2}$, M. Sarzi$^{1}$, N. Thatte$^{1}$, R. L. Davies$^{1}$, D. Krajnovi\'c$^{1}$ \\
$^{1}$University of Oxford, Denys Wilkinson Building, Keble Road, Oxford, OX1 3RH \\
$^{2}$University of Oxford, Rudolf Peierls Centre for Theoretical Physics, 1 Keble Road, Oxford, OX1 3RH}

\begin{document}

\date{29th September 2005}

\pagerange{\pageref{firstpage}--\pageref{lastpage}} \pubyear{2005}

\maketitle

\label{firstpage}

\begin{abstract}

  We present near infra-red (NIR) adaptive optics assisted
  spectroscopic observations of the CO ($\Delta\mu=2$) absorption
  bands towards the centre of the giant elliptical galaxy
  NGC~1399. The observations were made with NAOS-CONICA (ESO VLT) and
  have a FWHM resolution of 0\farcs15 (14pc).  Kinematic analysis of
  the observations reveals a decoupled core and strongly non-Gaussian
  line-of-sight velocity profiles (VPs) in the central 0.2 arcsec
  (19pc).  NIR imaging also indicates an asymmetric elongation of the
  central isophotes in the same region.

  We use spherical orbit-superposition models to interpret the
  kinematics, using a set of orthogonal ``eigenVPs'' that allow us to
  fit models directly to spectra.  The models require a central black
  hole of mass $1.2^{+0.5}_{-0.6}\times10^9M_\odot$, with a strongly
  tangentially biased orbit distribution in the inner 40pc.

\end{abstract}


\section{Introduction}

Super-massive black holes (SMBHs) are thought to be the only viable candidates for the massive dark object (MDO) observed at the centres of many nearby galaxies. Indeed, recent near infra-red (NIR) observations of the centre of the Milky Way have resolved individual stars orbiting in close proximity to the central MDO (which coincides with the radio source, Sgr A$^{*}$) and these data rule out all other plausible explanations for a MDO, other than a SMBH \citep{Schodel2003}.

Significantly, a relationship between the mass of the SMBH and the bulge luminosity of the host galaxy was discovered \citep{kor&rich95} and subsequently, a tighter correlation between the SMBH mass, M$_{\bullet}$, and the velocity dispersion of the bulge, $\sigma$ (the \MS relation) was measured \citep{f&m2000, geb2000} of the form
\begin{equation}
\label{m-s}
\log{\big{(}M_\bullet/M_{\odot}\big{)}} = \alpha + \beta\log(\sigma/\sigma_{0}).
\end{equation}
with $\sigma_{0}$=200\kms. \citet[hereafter T02]{T02} find $\alpha=8.13$ and $\beta=4.02$ and \citet[hereafter FF05]{FF05} find $\alpha=8.22$ and $\beta=4.86$. Similar relations with low scatter have also been found between M$_{\bullet}$ and the infrared luminosity L$_{\rm IR}$ \citep{m&h2003} and between $M_\bullet$ and bulge mass \citep{HR2004}.

It is believed that the mass accretion history of a SMBH is linked to the formation and evolution of its host \citep{H&K2000,zeeuw2003} and so such a precise relation, connecting quantities on vastly different scales, provides an important constraint on models of galaxy assembly. It would be particularly remarkable if it holds true for galaxies of different morphological types, which most likely underwent very different formation and evolution histories. For example, \citet{faber97} suggest that power-law ellipticals and spiral bulges formed dissipatively whereas core-like ellipticals formed from mergers, yet both appear to follow the same relation. In practice, the relation can also be used to measure the mass of a host galaxy black hole (BH) where a dynamical estimate is not possible \citep{AR2002,YT2002}.

\subsection{Contention}

\begin{center}
  \begin{figure}
    \centering
    \includegraphics[width=0.5\textwidth]{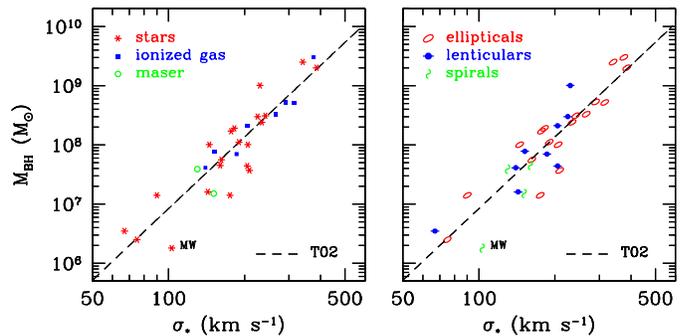}
    \caption{The sample of T02 and their best fit correlation of the form of (\ref{m-s}) with $\alpha=8.13$ and $\beta=4.02$. \textbf{Left:} the symbols indicate the technique used to derive the black hole mass. \textbf{Right:} the symbols represent the different morphological types.}
    \label{fig:T02}
  \end{figure}
\end{center}

There has been considerable debate over the values of the parameters $\alpha$ and $\beta$ \citep{f&m2000,geb2000,T02,FF05}. The slope $\beta$ is crucial for comparison with theoretical models that attempt to explain the \MS relation, but currently, the sample of galaxies used by T02 (Fig. \ref{fig:T02}) and FF05 is somewhat limited and biased. Of the 31 galaxies in the sample of T02, 18 are elliptical, 9 are lenticular and 4 are spiral (see Fig. \ref{fig:T02}). Although there are roughly equal numbers of power-law and core ellipticals, there remain few high dispersion and low dispersion galaxies where any deviations from the canonical slope will be most obvious. Such bias is not surprising considering that, until recently, only one telescope could perform dynamical mass estimates with the required spatial resolution: the Hubble Space Telescope (HST).

In order to reliably estimate the mass of the black hole, it is necessary to resolve kinematics in the region of space where the black hole potential dominates over the potential of the stars, a point stressed by FF05. The radius of this sphere of influence (SoI) is of order
\begin{equation}
\label{soi}
r\sim\frac{GM_{\bullet}}{\sigma^{2}_{\star}}
\end{equation}
where $G$ is the gravitational constant and $\sigma_{\star}$ is the average stellar velocity dispersion of the spheroidal component. Ground based observations have, in the past, been limited by atmospheric seeing to a resolution of around 1\arcsec or more. There are very few galaxies close enough and with sufficiently large SMBHs to have a SoI larger than this, so nearly all of the sample of T02 and FF05 comes from HST data (diffraction limit of $\sim$0\farcs04 at 4000\AA). However, HST has its limitations. 

Massive giant ellipticals which populate the high dispersion region of the \MS plane are rare (so on average more distant) and have low surface brightness centres with flat core-like density profiles \citep{faber97}. Although the total luminosity of such galaxies is high, they have too low a surface brightness to efficiently achieve high signal-to-noise (SNR) observations with HST's 2.4m primary. Conversely, high SNR nuclear observations of galaxies with steep power-law photometry \citep{faber97} are feasible due to their large numbers (therefore proximity) and rapid increase in surface brightness towards the nucleus. 
Hence, less massive ellipticals with steep power-law surface brightness profiles are common in the sample of T02, but very massive ellipticals with core-like surface brightness profiles are rare.

The low dispersion region of the \MS plane is populated by low mass bulges in spirals that tend to be dusty and obscured at the centre. Extinction is reduced at longer wavelengths (the extinction at $2.3 \mu$m is only 10\%\ of that at V-band), but HST is unable to probe the nuclei of such galaxies as it does not have a suitable IR spectrograph. Therefore, less massive dusty spirals are also under-represented in the current sample.

The outcome of these selection effects is an uneven distribution of galaxies along the $\sigma$ axis: 70\% of the T02 sample have $120$\kms$ < \sigma < 250$\kms. It is imperative that galaxies with high and low velocity dispersions be investigated, both to verify the uniformity of the \MS relation over different morphological types and a larger dispersion range, but also to better determine the nature of the relation. The slope of the \MS relation remains contentious (T02, FF05) because data points at the extremities, crucial for defining the slope, are sparse. Furthermore, some attractive theories of galaxy evolution predict departures from the current power-law relation which would only be detectable with more data at high and low dispersions \citep{H&K2000,Zhao02}.

Modern adaptive optics (AO) facilities on large 8m class ground based telescopes (diffraction limit of 0\farcs059 at 2.3\mum) are the key to solving this problem. The spatial resolution achieved with adaptive optics is limited primarily by the diffraction limit of the telescope. In the case of NIR observations with an 8m primary, this can match the spatial resolution of HST in the optical. Furthermore, 8m telescopes deliver high SNR observations of low surface brightness objects and the NIR spectrographs available on ground based telescopes are better able to probe obscured dusty regions. We have undertaken a careful study to identify targets with bright reference stars close to the galaxy centre which are under represented in the current \MS plane. 

\subsection{NGC 1399}

The first of these targets to be observed is the giant elliptical NGC 1399, which is the most luminous galaxy in Fornax and has a core-like surface brightness profile characteristic of a cD type galaxy \citep{kill&bick88,schom86}. Many dynamical studies in the visible have been undertaken in the past \citep{bick89,don95,f&i&h89,graham98,lon94,saglia00} although none have been made in the NIR or with the spatial resolution available with an AO system. 

The velocity dispersion of the spheroidal component $\sigma_{\star}$ is calculated in different ways by different authors \citep{f&m2000,geb2000}. As highlghted by T02, \citet{f&m2000} and subsequently FF05 calculate the RMS dispersion within a circular aperture of radius $r_e/8$ while \citet{geb2000} and T02 use the (luminosity weighted) RMS dispersion within a slit aperture of length $2r_e$. Whether or not the different definitions affect the measured slope of the relation, it is important to measure the corresponding values for NGC 1399 so like can be compared with like. 

The effective radius of NGC 1399 is reported to be 40\arcsec\ by
\citet{vau91}. Using the longslit data of \citet{graham98} to give a
slit aperture of length $2r_e$ centred on the galaxy ($\pm40$\arcsec)
and the photometry of \citet{Lauer05}, we calculate the luminosity
weighted RMS dispersion along the slit aperture to be $317\pm3$ \kms\
(the quoted error is an estimate of random error only). NGC 1399 is
therefore at the top of the T02 \MS plane, with three other galaxies
that anchor the relation (M87, IC 1459 and NGC 4649). The predicted BH
mass for NGC~1399 is $8.7\times10^8 M_{\odot}$ so the SoI would be
0\farcs34 (33pc); much smaller than even the best seeing at the best
observatory sites.

Following \citet{f&m2000}, the RMS dispersion within a circular
aperture of radius $r_e/8$ was estimated from the central
$2\arcsec\times5\arcsec$ of \citet{graham98}'s data to be
$329\pm4$\kms\ (as before, the quoted error is a measure of the random error).
The relation of FF05 then predicts a BH mass of $1.89\times10^9
M_{\odot}$ and the SoI would be 0\farcs78 (75pc): still not well
resolvable with the best seeing conditions.

Fortunately, NGC~1399 is ideally suited to AO assisted observations. A bright ($m_v=13.8$) reference star exists only 17\farcs6 away from the galaxy centre. The galaxy is almost spherical with little or no rotation, so the slit is free to be aligned to the galaxy centre and the AO reference star. This allows us to monitor the AO correction as a function of time.

One problem that plagues all stellar-dynamical estimates of black hole
masses is the degeneracy between mass and anisotropy
\citep{binney&mamon82}.  Consider the case of a
spherical galaxy.  Using the Jeans equation, the mass enclosed within
radius~$r$ can be written as \citep{kor&rich95}
\begin{eqnarray}
M(r)  &=&  {v^2r\over G}+
 {\sigma_{r}^{2} r \over G}
\left[ -{\dd \ln j \over \dd \ln r}
  -{\dd \ln \sigma_r^2 \over \dd \ln r}
  -\left(1-{\sigma_\theta^2\over\sigma_r^2}\right)\right.\cr
&&\qquad\qquad\qquad\qquad
  -\left.\left(1-{\sigma_\phi^2\over\sigma_r^2}\right)\right],
\label{eq:kr95}
\end{eqnarray}
where $v$ is the rotation velocity, $\sigma_r$, $\sigma_\theta$,
$\sigma_\phi$ are the radial and azimuthal components of the velocity
dispersion and $j$ is the deprojected luminosity density.  The
first two terms in square brackets can be estimated almost directly
from observations.  Both are positive for the vast majority of
galaxies.  The last two terms, however, depend on the the unknown
anisotropy and can take either sign.  Their effect on $M(r)$ is
minimised for galaxies with steep $j(r)$ profiles, steep velocity
dispersion profiles $\sigma_r$ and rapid rotation~$v\ne0$, all of
which tend to be satisfied in power-law galaxies.  Core galaxies like
NGC~1399, however, tend to be non-rotating with shallow density and
dispersion profiles.  For such galaxies it is particularly important
to constrain the anisotropy by modelling at the detailed shape of the
galaxy's line-of-sight velocity profiles (VPs), for which high
signal-to-noise spectra are essential \cite{Gerhard93}.

Using the NAOS AO system coupled with the CONICA NIR imager / spectrograph at the European Southern Observatory's Very Large Telescope (ESO VLT) we have resolved the SoI of NGC 1399 and measured its stellar kinematics using the CO absorption bands at 2.3\mum\ and the CaI absorption feature at 2.26\mum. The SNR of the spectra range from $\sim90$ to $\sim20$ over the region used to extract kinematics, with a SNR of $\sim70$ at the CO bandhead (2.3\mum). We use these kinematics to construct a spherical orbit superposition model for the galaxy to estimate the mass of the central MDO. Throughout this paper we assume a distance of 19.9Mpc to NGC 1399 \citep{tonry2001}; the reader is reminded that BH mass scales linearly with assumed distance.

The structure of this paper is as follows. The data reduction techniques are discussed in Section \ref{sec:reduction}; the kinematic analysis is discussed in Section \ref{exkin}; the imaging and kinematics are presented in Section \ref{sec:results} and the discussion of their implications is contained in Section \ref{sec:datadiscus}. The dynamical modelling is described and discussed in Section \ref{sec:modelling}. Finally, Section \ref{sec:conc} concludes.

\section[Data and Reduction]{Data and Reduction}
\label{sec:reduction}
\subsection{Observations}
AO assisted K band images (Ks filter) and K band long-slit spectra (SK filter) of the nuclear region of NGC 1399 were obtained with NAOS-CONICA \citep{naos, conica} on the nights of 30/11/03 and 01/12/03. The spectral range extended from 1.79\mum--2.45\mum, with a scale of 0.972nm per pixel, although atmospheric transmission limits high SNR data to 1.95\mum--2.45\mum. The spatial scale of the spectroscopy was 0\farcs0543 per pixel; the scale of the imaging was 0\farcs027 per pixel. The slit width was 0\farcs172 (17pc) corresponding to an instrumental resolution ($\lambda/\Delta\lambda$) of 880 at 2.3\mum~(as measured from the width of the arc lines) and an instrumental broadening of $\sigma_{inst}=145$\kms. The conditions while observing were excellent with the seeing varying between 0\farcs4 and 0\farcs6.

A total of 27000 seconds (7.5 hours) of useful on-source spectroscopic integration was achieved with individual exposures lasting 300 seconds each. Airmass ranged from 1.0 to 1.4, although 85\% of exposures were made with airmass $<1.2$. The standard ABBA technique 
of nodding back and forth along the slit removed the need for separate sky exposures and Fowler readout mode was used to minimise readout noise. AO assisted images of NGC 1399 were also taken for a total of 80 seconds with the Ks filter together with an equivalent number of sky exposures.

In order to remove the complex atmospheric transmission curve (referred to as telluric absorption) from the object spectra, it was necessary to observe several \emph{telluric standard stars} (see Sec. \ref{telcor}) which have almost intrinsically featureless spectra in the NIR (e.g. very hot O or B stars). As the strength of telluric absorption depends on the airmass, on the first night telluric standards were observed at airmasses of 1.02 (HD25631) and 1.40 (HD25631), and on the second night telluric standards were observed at airmasses of 1.12 (HD480) and 1.25 (HD41814). The difference in airmass between telluric and galaxy exposures on each night was never more than 0.2 airmasses. Two kinematic template stars were also observed at low airmass: HD11931 (K4III) and HD25840 (M0III).

All stars (telluric standards and kinematic templates) were observed with an identical spectrograph configuration to the galaxy observations. However, for the stellar observations alone, it was necessary to reduce the higher order gain of AO correction to ensure that the FWHM of the point spread function (PSF) was larger than the width of the slit to match the spectral resolution of the galaxy and stellar observations. With full AO correction, the FWHM of the PSF was $\sim$0\farcs1 which would have been significantly smaller than the 0\farcs172 wide slit.

The position angle (PA) of the slit was $5.06^\circ$ so as to include the galaxy centre and AO reference star in the slit. This allowed us to monitor and asses the AO correction. We assume that the SMBH lies at the most luminous region of the galaxy (also assumed to be at the centre of the galaxy) so it is important to position the slit to sample this region or to be able to quantify any offset from it. Prior to acquiring the galaxy, an image of the slit on the detector was made by removing the grism and illuminating the slit with the flat field lamp. The slit image was then used as a bias for the subsequent acquisition images to precisely align the slit. Special care was taken with the first acquisition of the first night to ensure the PA of the slit intersected the AO reference star and the brightest part of the galaxy. The PA was subsequently held fixed for all observations and only shifts perpendicular to the slit length (i.e. along the slit's minor axis) were made to maintain the slit position on the star and the galaxy centre. On the second night, the same PA was verified to hold the star and the galaxy centre in the slit and then shifts were made along the slit's minor axis as before.

\subsection{Reduction Techniques}

The data reduction was completed with the aid of the IRAF\footnote{IRAF is distributed by National Optical Astronomy Observatories (NOAO), http://iraf.noao.edu} and ECLIPSE\footnote{ECLIPSE is a reduction package developed by the European Southern Observatories (ESO), http://www.eso.org/projects/aot/eclipse/} packages as well as custom IDL\footnote{Interactive Data Language, Research Systems, Inc.} scripts, incorporating use of the IDL Astronomy User's Library \citep{landsman93}\footnote{http://idlastro.gsfc.nasa.gov/}. 

The spectroscopic data were initially reduced following the standard ABBA technique for NIR data reduction\footnote{For further information on NIR data reduction, the reader is referred to the ISAAC data reduction guide (the NACO data reduction guide is still being constructed)} which has many advantages: the timescale on which the background subtraction is achieved is as short as possible, helping to correct for the variability of the NIR sky; any residual sky in a single A-B frame cancels with the residual in the B-A frame, assuming a uniform sky field; the pixel-to-pixel subtraction is very well suited to removing systematic errors. However, such a technique does not optimise the random noise (sky, thermal, dark, readout) as a background exposure of the same duration as the object exposure is subtracted from each pixel. 

In an effort to increase the SNR, the background level of each pixel was interpolated as a function of time from all the data frames. Pixels with significant source (galaxy) counts were excluded when fitting a 3rd order polynomial as a function of time to each pixel position. In order to gauge the change in the SNR from interpolation, the random noise of the sky dominated region between galaxy and the AO reference star was measured as a function of wavelength and compared to the noise of the frames without background interpolation: the interpolated background showed a significant decrease in random noise ($\sim\sqrt{2}$ lower). This led to a significant increase in the final SNR of the data while maintaining the pixel-to-pixel subtraction to reduce systematics.
However, the residuals from one object-sky pair do not necessarily cancel with the next pair, unlike the classical reduction technique for ABBA sequence observations (assuming a uniform illumination). The interpolation is also susceptible to bias from bad pixels which can propagate into neighbouring frames. To counter this problem, the frames were cleaned of bad pixels prior to (object and sky frames), during (sky pixels only) and after the interpolation (background subtracted pixels). Indeed, a consequence of interpolating the background was the identification of faulty pixels on the detector and the construction of an very accurate bad pixel map.

The timing and conditions of the spectroscopic data make it well suited to sky interpolation: 7.5 hours of observations were performed on neighbouring nights with limited interruption between the exposures on each night and the atmospheric conditions during the two nights were excellent and stable.

After correction for the odd-even effect of the detector, sky subtraction, flat fielding, bad pixel correction, field transformation and wavelength calibration, the exposures were aligned along the spatial axis (to an accuracy of a few tenths of a pixel) by using the centroid of a Gaussian fitted to the light profile of the AO reference star. Reduction of telluric standard star spectra followed the same sequence except with only standard ABBA sky subtraction.

The image data was reduced in the standard manner for NIR observations. After correcting for odd-even effect, the sky exposures were subtracted from the object frames, followed by flat fielding. The centroid of a Gaussian, fitted to the AO reference star, was then used to align the individual exposures to a tenth of a pixel accuracy. 

\subsection{Telluric Correction}
\label{telcor}
Molecular gas in the Earth's atmosphere (CO$_2$, H$_2$O etc.) produces an absorption spectrum known as telluric absorption and this must be removed from astronomical observations. The galaxy and kinematic template observations were corrected for telluric absorption in the standard manner \citep{omo93}, the details and caveats of which are described below. 

The stellar continuum must first be removed from the telluric standard star spectrum. We therefore divide through by a blackbody spectrum of appropriate temperature to the star. This is a blind division; one cannot fit the temperature as the atmospheric transmission curve is still imprinted on the spectrum (in addition to any instrumental response). 

Telluric absorption varies with airmass so the strength of the sharp prominent absorption features in the normalised telluric spectrum must be matched to those in the object spectrum (of similar airmass); division of this optimised standard then removes the prominent (high frequency) features and, we assume, the low frequency component of the atmospheric transmission curve. The telluric spectrum, now free from stellar continuum, is not yet normalised so we divide by a linear fit to regions where we \emph{believe} the atmospheric transmission to be $\simeq$100\%: (2.035-2.04)\mum, (2.09-2.15)\mum and (2.21-2.22)\mum. The increased strength and occurance of telluric features beyond 2.3\mum~causes atmospheric transmission to be consistently below unity, and falling with increasing wavelength so one must extrapolate the continuum from shorter wavelengths. The absorption depth in the now normalised telluric spectrum, $A(\lambda)$ is then amended using the expression
\begin{equation}
\label{eq:tel}
A'(\lambda) = 1.0 - F \big{[}1.0 - A(\lambda) \big{]}
\end{equation}
where $F$ is a free parameter, to account for small variations in airmass between the object and telluric standard star observations. Note that this linear correction for airmass is only an approximation to a more complicated response and is only effective for very small differences in airmass. The initial normalisation was then removed by multiplying back by the original linear fit. Every telluric spectrum also required a small shift in wavelength (around a few tenths of a pixel) to compensate for differences in wavelength zero-point between the object and telluric spectra. The optimisation of airmass and wavelength was initially automated by minimising residuals around the strong telluric features at (2.0 - 2.1)\mum~in the telluric divided object spectrum. Fine adjustments were then made to further optimise the removal of prominent telluric absorption features at the region of interest (the CO band-heads after 2.3\mum). Note that in practice, while one can correct individual stellar exposures (due to the high SNR in each frame), galaxy exposures must be coadded in groups of similar airmass to increase the SNR before an accurate telluric correction can be determined.

Telluric correction is not without complication though. The telluric correction is optimised for the removal of sharp, prominent, high frequency absorption features. Accordingly, there is likely to be minimal residuals from such in the corrected spectra. However, variation in the continuum normalisation of a galaxy spectrum is known to introduce systematic effects into the derived kinematics \citep{vdMetal94}. Due to the blind division of a blackbody spectrum and the lack of continuum reference in the telluric spectra after 2.3\mum, error in telluric correction will most likely manifest itself as error in the continuum normalisation of the telluric spectrum after the application of (\ref{eq:tel}). This would propagate into an error in the continuum level of the object spectrum (galaxy or kinematic template). Such an error may not be uniform over the length of the spectrum: there is likely to be a higher chance of error where the normalisation was completely extrapolated, after 2.3\mum. To help compensate for such systematic effects, we include a polynomial continuum correction when extracting kinematics (see Sec. \ref{exkin}), but the constraint for the correction is the minimisation of $\chi_{\rm s}^2$ (\ref{eq:chisqspec}), which does not guarantee to choose a solution free of systematics. Furthermore, this continuum correction is additive whereas any real error would be divided into the object spectrum. 

Different functions were used to normalise the galaxy and kinematic template spectra after telluric correction. The galaxy continuum appeared linear with wavelength and so at every position along the slit, a linear fit to the continuum shortward of the CO bands was sufficient. The continuum of each kinematic template shortward of the CO bandhead was clearly non-linear but was well fitted (and removed) by a blackbody spectrum. As the continuum after 2.3\mum\ must be extrapolated, it was necessary to fit slowly varying functions with relatively little freedom. Note that with all continuum fits, care was taken not to include obvious absorption or emission features, or areas of significant telluric absorption.

\subsection{Kinematic Templates}

The choice of kinematic template(s) is important to accurately extract kinematics from the galaxy spectra. The systematic errors introduced into kinematics by use of a poor template are well studied \citep{vdMetal94}, although it is difficult to numerically quantify such effects. At best, we can say that different templates appear to introduce systematic offsets into the VP parameters. Ideally, a large library of spectral types should be available so that an optimal mix can be found, but in this case, due to time constraints, only two templates were observed with the same instrumental setup as the NGC 1399 observations. However, it is possible to check if the templates are well matched to the luminosity weighted population of the galaxy.

The CO (2--0) band head at 2.2935\mum~is an indicator of stellar type: a linear relation has been found between the equivalent width (EW) of the first CO feature, W$_{{\rm CO (2-0)}}$ and the stellar type \citep{k&h86,omo93}. Although this relation is based on observations of individual stars, it can be extended to galaxy populations by applying a correction based on the velocity dispersion of the system \citep{oliva95,ttg2000}. The relationship differs between giants and supergiants but this should not be a problem for the old, giant dominated, population of an elliptical galaxy such as NGC 1399. To estimate the galaxy EW, the (rest frame) wavelength range over which the CO (2-0) EW is defined must be shifted to the velocity frame of the galaxy and the EW measurement must be corrected for the galaxy's velocity dispersion. The kinematic properties of the galaxy are not known a priori and depend on the template used to extract them. However, it is possible to account for this and calculate reasonable limits on the EW of the galaxy.

Such analysis was performed for NGC 1399. The variation of EW with velocity dispersion was simulated using the kinematic templates and the necessary quadratic correction (in $\sigma$) found; this was then applied to the galaxy EW measurements. Furthermore, all measurements were scaled by a constant factor (the correction of \citet{oliva95}) to correct for the instrumental resolution of CONICA, allowing direct comparison with the relation of \citet{omo93}. The results are presented in Fig. \ref{fig:paramresults} and discussed in Sec. \ref{sec:results}.

\subsection{AO correction and PSFs}
\label{sec:AO}

\begin{center}
  \begin{figure}
  \includegraphics[width=0.45\textwidth]{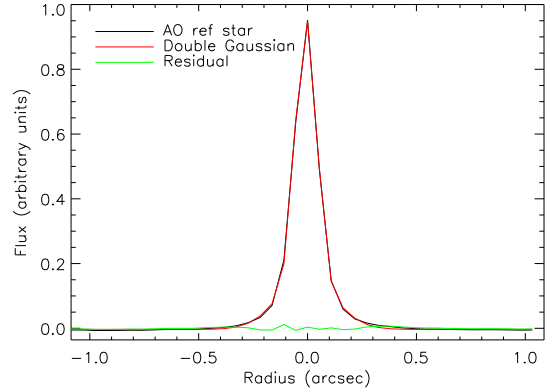}
  \caption{The profile of the AO reference star together with a double Gaussian fit and the residuals from this fit. The best-fit $\sigma$ for each of the Gaussians are 0\farcs048 and 0\farcs122 and the total fluxes ($\gamma$s) are in the ratio 2:1 respectively. This 1D slit profile of the AO reference star is used to estimate the 2D on- and off-source PSFs.}
    \label{fig:psfs}
  \end{figure}
\end{center}

The quality of the AO correction can be estimated from the reference star, which was observed in the slit simultaneously with the galaxy. The 1D profile of the star (calculated from summing the flux over the same wavelength range that the kinematics are extracted from) is well fit by a double Gaussian as shown in Fig. \ref{fig:psfs}.

However, the correction and PSF vary further away from the reference star. Although the exact `off-source' PSF at the centre of the galaxy is unknown, it can be estimated. Using the NAOS preparation software v1.74\footnote{http://www.eso.org/observing/etc/naosps/doc/NAOS-PS-tool.html}, one is able to simulate how the PSF varies with seeing, airmass and distance from the reference star. However, the simulated PSFs do not account for slit effects, non-perfect data reduction (such as error in the field transformation or frame alignment) or drift of the the tip-tilt correction during long exposures. All these effects will further broaden the PSF. To account for such, we convolve simulated 2D on-source PSFs (with various airmass and seeing conditions) with a $\theta\times3.2$ top hat (3.2 pixels is the slit width and $\theta$ accounts for imperfect data reduction and tip-tilt drift). We optimise $\theta$ and the atmospheric conditions by minimising the difference between the 1D profile along $y=0$ and the observed 1D profile of the AO reference star (Fig. \ref{fig:psfs}). Good agreement between the simulated and observed PSFs is found for 0\farcs6 seeing, an airmass of 1.2 and $\theta=2.2$ pixels. To estimate the off-source PSF at the galaxy centre, we simulate the 2D off-source PSF with identical atmospheric conditions and convolve it with the same 2D kernel (we assume that the broadening effects are uniform over the field of fiew). The resulting on- and off-source PSFs are both well fitted by a double 2D Gaussian
\begin{eqnarray}
\label{eq:2ddoublegauss}
PSF(x,y) = & {\gamma_1\over2\pi\sigma_{x1}\sigma_{y1}} \exp\left[ -{1\over2} \left({x^2\over\sigma_{x1}}\right)^2 + \left({y^2\over\sigma_{y1}}\right)^2\right] \cr
+ &{\gamma_2\over2\pi\sigma_{x2}\sigma_{y2}} \exp\left[ -{1\over2} \left({x^2\over\sigma_{x2}}\right)^2 + \left({y^2\over\sigma_{y2}}\right)^2\right].
\end{eqnarray}
The best-fit parameters are given in Table \ref{tab:psfs} and are used to describe the 2D PSF in the dynamical modelling. The Strehl ratios also quoted in Table \ref{tab:psfs} are calculated from the unbroadened (pre-convolution) simulated PSFs: it makes no sense to compare the ideal Airy disk pattern with the slit-convolved PSF of the spectrograph. The FWHM of the off-source PSF along the x-axis is 0\farcs15 (2.75 pixels), corresponding to 14pc, which we adopt as our formal resolution. 

\begin{table}
\begin{tabular}{ccc}
\bf{Parameter} & \bf{On-source} & \bf{Off-source} \\\hline
$\gamma_1$     &   0.141  &  0.135 \\
$\sigma_{x1}$  &   0.956  &  1.071 \\
$\sigma_{y1}$  &   1.214  &  1.280 \\
$\gamma_2$     &   0.010  &  0.129 \\
$\sigma_{x2}$  &   3.667  &  3.501 \\
$\sigma_{y2}$  &   2.784  &  2.630 \\

FWHM along x (\arcsec)    & 0.135 & 0.150 \\
FWHM along y (\arcsec)    & 0.181 & 0.189 \\

Strehl Ratio (\%) &   38  &   30   \\

\end{tabular}
\caption{The characteristics of the on- and off-source PSFs: the best-fit parameters $(\gamma_1,\sigma_{x1},\sigma_{y1},\gamma_2,\sigma_{x2},\sigma_{y2}$) of a 2D double Gaussian (\ref{eq:2ddoublegauss}) fit to the PSFs after incorportaing broadening effects (illustrated in Fig.\ref{fig:psfs} and subsequently used in dynamical modelling); the FWHM of both broadened simulated PSFs along the x and y axes; the Strehl ratios of the simulated PSFs prior to incorporating additional broadening.} 
\label{tab:psfs}
\end{table}

The effect of contamination from the seeing-limited halo on the spectra and the knock-on effect on the VPs is not known precisely. According to (\ref{eq:chisqspec}), if the galaxy spectrum is a weighted sum of many different spectra with many different VPs, the problem is linear and $\chi_{\rm S}^2$ will be minimised for a similarly weighted mean VP. The galaxy spectra and kinematics are therefore only expected to be `diluted' by this effect. However, the dynamical modelling uses the estimate of the off-source PSF to \emph{fully} account for the shape of the PSF; the peculiar shape of the AO corrected PSF does not bias the models or the derived BH mass.

The quality of the PSF was assessed as a function of time from the AO reference star in the slit. The quality of the overall AO correction for the final coadded galaxy data can be marginally improved with frame selection. However, the cost in SNR was too high for any significant improvement in this case, so frame selection was not implemented. 

\section {Kinematics}
\label{exkin}
We assume that each galaxy spectrum~$G(\lambda)$ is a
Doppler-broadened version of some underlying ``average'' stellar
template~$T(\lambda)$.  If we could completely remove all continuum
features from $G$ and $T$, then we would have that
\begin{equation}
  \label{gal_conv}
  G(\lambda) = T(\lambda) \otimes L(v),
\end{equation}
where $L(v)$ is the unknown line-of-sight stellar velocity profile
and $\otimes$ denotes convolution.
In reality, it is impossible to completely remove all continuum
contamination from observed spectra by eye, nor do we know the correct
template~$T$ to use.  Furthermore, both $G$ and~$T$ are measured with
finite signal-to-noise, making it impossible to use a
simple deconvolution method to obtain~$L(v)$ directly
from~(\ref{gal_conv}).

Many methods have been proposed to address these problems.  We follow
\citet{rix&white92}, \citet{saha&will94} and \citet{vdM94} and extract
$L(v)$ using a direct pixel-fitting method. 
Taking the continuum-divided star and galaxy spectra we find $L(v)$
and its associated uncertainties by minimising
\begin{equation}
  \label{eq:chisqspec}
  \chi^2_{\rm s} \equiv \sum_{i=1}^{N_{\rm p}} \left[
 G_i - k\left(T \otimes L\right)_i - \sum_{l=0}^{2}{c_l (\ln\lambda_i)^l)}
    \over\Delta G_i \right ]^2,
\end{equation}
where $\Delta G_i$ is the measurement error in the $i$th galaxy
spectrum pixel, $N_{\rm p}$ is the number of pixels being fitted and
the third term corrects for any continuum that escapes our initial
continuum division.  The parameter $k\simeq1$ accounts for differences
in normalisation between $G$ and~$T$: we want our VPs to be
normalised with $\int L(v)\,\d v=1$.  We further assume that
$T(\lambda)$ is well approximated by a weighted average of known
stellar templates. As we only have two stellar templates available for
kinematic extraction, we define our optimal template to be
\begin{equation}
  \label{eq:optitemp}
  T(\lambda) = f T_1(\lambda) + (1-f) T_2(\lambda)
\end{equation}
where $f$ defines the relative fraction of each of the two available
templates $(T_1, T_2)$.

Unlike Fourier methods
\citep{sargent78,r&s72,f&i&h89,Bender90,vdM&Franx93}, pixel fitting
does not require any assumptions about window functions and allows us
to propagate the pixel-to-pixel error estimates in our measured
spectra directly into uncertainties in our LOSVDs.  Note that the NaI
doublet at 2.21\mum\ does \emph{not} appear to be well fit by our
templates (the line strength in the galaxy spectra is much higher than
that in our templates).  Thus the fitting range for extraction of
kinematics is from 2.249\mum\ to 2.438\mum\ (rest frame) to include the
CaI feature and the CO bands.  We use two different parametrisations
for $L(v)$, each of which is discussed separately below.

\subsection{Gauss--Hermite parametrisation of the VP}
\label{sec:ghvps}
VPs are expected to be reasonably close to Gaussian.
A convenient way of parametrising a VP~$L(v)$ is by using a truncated
Gauss--Hermite expansion
\citep{Gerhard93,vdM&Franx93},
\begin{equation}
   \label{eq:GHseries}
\Lgh(v) = {\gamma\over\sqrt{2\pi}\sigma}
  \exp\left[-{1\over2}\left(v-V\over\sigma\right)^2\right]
  \sum_{i=0}^N h_iH_i\left(v-V\over\sigma\right),
\end{equation}
in which one starts from a Gaussian with scale factor~$\gamma$,
mean~$V$ and dispersion~$\sigma$ and uses a weighted sum of Hermite
polynomials~$H_i$ to quantify deviations of $L(v)$ from this
underlying Gaussian. This parametrisation is degenerate: there is a different
set of coefficients $\{h_i\}$ for each choice of $(\gamma,V,\sigma)$
and a number of different procedures have been used to fit spectra
using~(\ref{eq:GHseries}).  Before describing our own procedure, we
review the motivation for these earlier methods.

\subsubsection{Perfect data}
\label{sec:idealghvps}
Consider first an idealised situation in which we knew $L(v)$ perfectly.
Then there is a unique set of expansion coefficients $\{h_i\}$ for any
(sensible) choice of Gaussian $(\gamma,V,\sigma)$.  To see this,
recall that the $H_i$ satisfy the orthogonality relation \citep{vdM&Franx93},
\begin{equation}
{1\over2\pi}\int_{-\infty}^\infty \exp(-x^2)H_i(x)H_j(x)\,\d x =
{1\over2\sqrt\pi} \delta_{ij},
\end{equation}
from which it is easy to see that choosing
\begin{equation}
\label{eq:GHcoeff}
h_i = {1\over\sqrt2\gamma}
\int_{-\infty}^\infty \exp\left[-{1\over2}\left(v-V\over\sigma\right)^2\right]
L(v)H_i\left(v-V\over\sigma\right)\,{\rm d}v
\end{equation}
minimises the mean-square deviation
\begin{equation}
\label{eq:chisqzero}
\chi^2_0 \equiv \int_{-\infty}^\infty [L(v)-\Lgh(v)]^2\,\d v
\end{equation}
of the expansion~(\ref{eq:GHseries}) from the VP~$L(v)$.  Since the
$H_i$ form a complete set, we can make $\chi^2_0$ arbitrarily small
for smooth $L(v)$ simply by increasing the number of terms~$N$
included in the series, although the expansion~(\ref{eq:GHseries}) is
of course useful in practice only if it can provide a good fit for
small~$N$.

We note that when minimising $\chi_0^2$ at fixed
$(\gamma,V,\sigma)$:
\begin{enumerate}
\item the Gauss--Hermite coefficients $h_i$ are modified
  moments~(\ref{eq:GHcoeff}) of~$L(v)$;
\item the Hessian $\p^2\chi^2_0/\p h_i\p h_j$ is diagonal, meaning that the
  $h_i$ are independent;
\item the $h_i$ are also independent of the choice of~$N$;
\item if we choose $(\gamma,V,\sigma)$ to be
  the parameters of the Gaussian that minimises~(\ref{eq:chisqzero}) then 
$(h_0,h_1,h_2)=(1,0,0)$ and $(h_3,h_4)$ measure
the lowest order anti-symmetric and symmetric deviations from this
Gaussian.
\end{enumerate}
Because of this last point, most Gauss--Hermite parametrisations of
VPs fix $(h_0,h_1,h_2)=(1,0,0)$ and use
$(\gamma,V,\sigma,h_3,h_4,\ldots,h_N)$ as the free parameters in the
fit.  For our purposes, this approach has the drawback of introducing
nontrivial correlations between the Gaussian parameters
$(\gamma,V,\sigma)$ and the $\{h_i\}$, making these parameters
cumbersome to use when comparing to dynamical models.

\subsubsection{Real data}\label{sec:realghvps}
In reality we do not have direct access to $L(v)$.  Instead we
constrain it by investigating how well a parametrised form, such
as~(\ref{eq:GHseries}), affects the fit to the discretely sampled
galaxy spectrum~$G_i$ using the $\chi^2_{\rm s}$ given
by~(\ref{eq:chisqspec}).  As \citet{vdM&Franx93} point out,
eq.~(\ref{eq:chisqspec}) reduces to~(\ref{eq:chisqzero}) in the limit
of high resolution, finely sampled spectra and sharp template
features.  Here we consider the case where this limit does not apply.
Let us assume that the continuum has been perfectly removed from $G_i$
and that $T$ is the correct stellar template.
Then~(\ref{eq:chisqspec}) becomes
\begin{equation}
\label{eq:chisqspectwo}
\chi^2_{\rm s} = \sum_i \left[ (T\otimes(L-\Lgh))_i+n_i\over\Delta G_i\right]^2,
\end{equation}
where $L(v)$ is the galaxy's real underlying VP and $n_i\equiv
G_i-(T\otimes L)_i$ is the noise in the $i^{\rm th}$ pixel.  Now if we fix
$(\gamma,V,\sigma)$ and minimise~(\ref{eq:chisqspectwo}) with respect to
the $\{h_i\}$, then:
\begin{enumerate}
\item in the absence of noise, the coefficients $h_i$ are still the modified
  moments~(\ref{eq:GHcoeff}) of $L(v)$;
\item the $h_i$ are not independent since the Hessian $\p^2\chi^2_{\rm
    s}/\p
  h_i\p h_j$ is no longer diagonal;
\item since the $h_i$ are not independent, in the presence of
  noise there is a different set of
  $h_i$ for each choice of $N$;
\item if we choose $(\gamma,V,\sigma)$ to be the parameters of the
  best-fit Gaussian to~(\ref{eq:chisqspec}), then the minimum
  $\chi^2_{\rm s}$
  will not occur at precisely $(h_0,h_1,h_2)=(1,0,0)$.
\end{enumerate}

Our procedure for fitting Gauss-Hermite coefficients is motivated by
these points and by our desire to have a set of parameters that depend
linearly on $L(v)$.  The procedure is as follows:
\begin{enumerate}
\item Choose $(\gamma,V,\sigma)$ to be the parameters of the best-fit
  Gaussian to the VP: find $(\gamma,V,\sigma)$, template
  fraction~$f$ and continuum parameters~$c_l$ that
  minimise~(\ref{eq:chisqspec}) with $k=h_0=1$ and $h_1=h_2=\cdots=0$.
\item Holding $(\gamma,V,\sigma)$ fixed at their best-fit values, we find
  the $h_i$, $c_l$ and $f$ that minimise~(\ref{eq:chisqspec}).  Having
  found these~$h_i$, the normalisation is
  $k=\gamma(h_0+h_2/\sqrt2+h_4\sqrt{3/8}+\cdots)$.
\item Finally, we divide $\gamma$ and the $h_i$ by $k$.
\end{enumerate}

We use a standard Levenberg-Marquardt routine to carry out the
minimisations in the first two steps.  Our best-fit parameters are
$(h_0,\ldots,h_N)$, along with their covariances and the choice of
$(\gamma,V,\sigma)$.  Note that there are no errors associated with
$(\gamma,V,\sigma)$ in our version of the Gauss--Hermite
parametrisation: they merely reflect the Gaussian around which we have
chosen to expand~$L(v)$.  Choosing the best-fit Gaussian here lets us
make a straightforward comparison of our kinematics with earlier work.
In practice we find that our method yields $(h_0,h_1,h_2)$ that differ
from $(1,0,0)$ by around $(0.1,0.04,0.04)$, but with very strongly coupled
errors among all the even~$h_i$.  We describe how we deal with these
covariances in section~\ref{sec:modelfit} below.

We have tried using standard simultaneous $N+1$ parameter fits
to $(\gamma,V,\sigma,h_3,\ldots,h_N)$ \citep{vdM&Franx93}, but our VPs are so strongly
non-Gaussian at the centre of the galaxy that the usual linear approximations among the errors in
these parameters \citep{vdM&Franx93} break down and we find multiple minima in $\chi_{\rm S}^2$ (\ref{eq:chisqspec}).  For spectra at (-0\farcs08,0\farcs02) this process can yield $h_4>0.6$ and alarmingly low values
of~$\sigma$ ($\sim 250$\kms) which describe a triple peaked VP. In fact, for such values of $h_4$ it can be shown that even the idealised~$\chi^2_0$ (\ref{eq:chisqzero}) has multiple minima in~$\sigma$ and $h_4$.

\subsection{VP histograms}
\label{sec:histvps}
One might expect that the putative BH in NGC~1399 would cause
high-velocity wings in the central VPs, which might not be captured
well by the low-order Gauss--Hermite parametrisation above.  Therefore
we also fit ``non-parametric'' VPs, in which we choose $N$ regularly
spaced velocities $v_1<v_2<\cdots <v_N$ and parametrise $L(v)$ as
the histogram
\begin{equation}
\Lhist(v) = \sum_{i=1}^{n_v} L_iS_i(v),
\end{equation}
where the step function~$S_i(v)=1$ if $v_i<v<v_{i+1}$ and is zero
otherwise.

Given parameters $L_1,\ldots,L_{n_v}$ it is straightforward to calculate
the convolution~(\ref{gal_conv}) of this $\Lhist(v)$ with a stellar
template.  For any given galaxy spectrum~$G$ there will be many sets
of parameters that produce good fits to the spectrum, but most of them
will be unrealistically jagged.  Therefore, instead of minimising the
$\chi^2_{\rm s}$ given by eq.~(\ref{eq:chisqspec}) directly, we minimise the
penalised $\chi^2_{\rm p}=\chi^2_{\rm s}+P[L_i]$, where the penalty function
\begin{equation}
P[L_i] = \alpha\sum_i (L_{i+1}-2L_i+L_{i-1})^2
\label{eq:penaltyfn}
\end{equation}
uses the mean-square second derivative of $L(v)$ as a measure of the
jaggedness of the solution.

Our procedure for fitting~$L_i$ is simple:
\begin{enumerate}
\item Find the best-fit smooth~$L_i$, continuum parameters~$c_l$
  and template fraction~$f$ by
  minimising the penalised $\chi^2_{\rm p}=\chi^2_{\rm s}+P[L_i]$ with
  $k=1$.
\item Set the normalisation factor to $k^{-1}=\sum_i
  (v_{i+1}-v_{i})L_i$ and rescale the~$L_i$.
\end{enumerate}

This makes no attempt to impose the obvious non-negativity constraint
on the~$L_i$.  While the resulting VP histograms are fine for
``by-eye'' comparisons of one VP against another, they are not
suitable for direct comparison against models.  So, in
\S\ref{sec:modelfit} we describe a variation on this fitting procedure
that takes account of the correlations among the~$L_i$ and removes the
bias introduced by the penalty function.

Based on an average separation of 300\AA~between the CO bands in the
fitting range ($^{12}$CO$(2-0)$, $^{12}$CO$(3-1)$, $^{12}$CO$(4-2)$,
$^{12}$CO$(5-3)$, $^{12}$CO$(6-4)$), the maximum relative velocity we
can reasonably hope to measure is $\sim 1900$\kms.  The systematic
velocity of NGC~1399 is $\sim1500$\kms\ so we divide each LOSVD into
$n=50$ equispaced velocity points between $v_1=-1000$\kms and
$v_n=4000$\kms. We choose $\alpha=4\times10^7$, which is the minimum
value required to give smooth non-parametric VPs consistent with our
outer Gauss-Hermite VPs.


\begin{figure*}
 \begin{minipage}[c]{\textwidth} 
   \includegraphics[width=\textwidth]{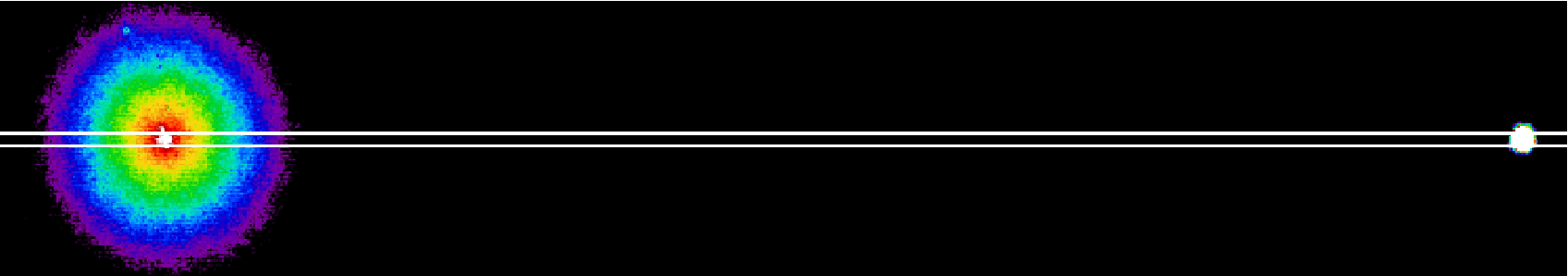}
 \end{minipage}
 \begin{minipage}[c]{\textwidth} 
   \includegraphics[width=\textwidth]{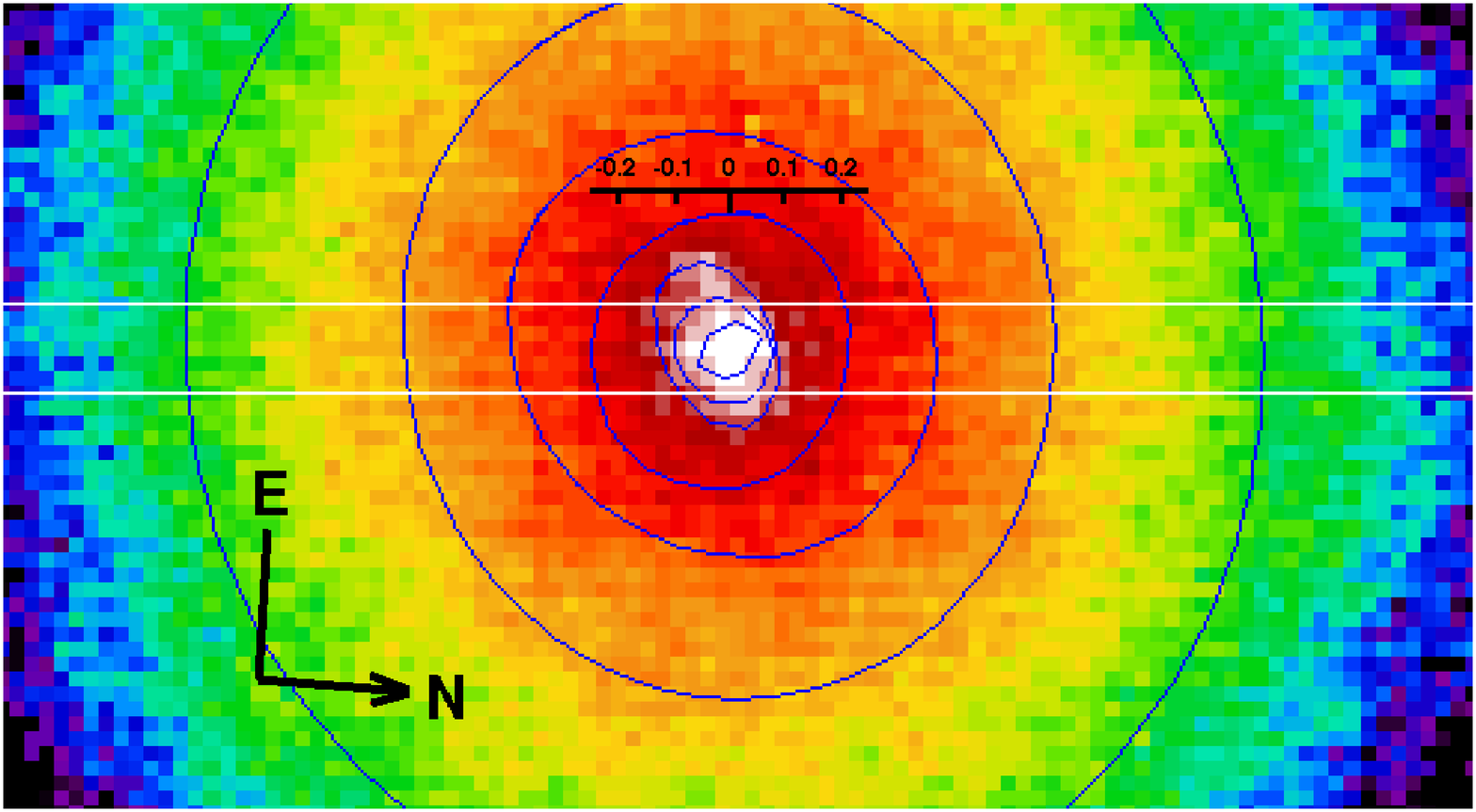}
 \end{minipage}
  \caption{\emph{Top}: A Ks band image of NGC 1399 showing the slit alignment (white) on the galaxy and the AO reference star (saturated on this colormap); the PA of the slit is $5.06^\circ$ so that the reference star is approximately due north of the galaxy nucleus; the nucleus of the galaxy and the star are separated by 17\farcs5. \emph{Bottom}: the same as the above but magnified and centred on the nucleus of NGC 1399 with isophote ellipses over plotted in blue. Note the elongation of the nucleus to the south-east. Each pixel is 27mas wide (the pixel size of the spectroscopic data was twice this at 54mas). The angular scale is given in arcseconds.}
 \label{fig:1399image}
\end{figure*}

\begin{center}
  \begin{figure}

    \begin{minipage}[c]{0.03\textwidth}
      \begin{flushleft}
	\begin{sideways}{}Flux\end{sideways}
      \end{flushleft}
    \end{minipage}%
    \begin{minipage}[c]{0.45\textwidth} 
      \begin{flushright}
	\includegraphics[width=\textwidth]{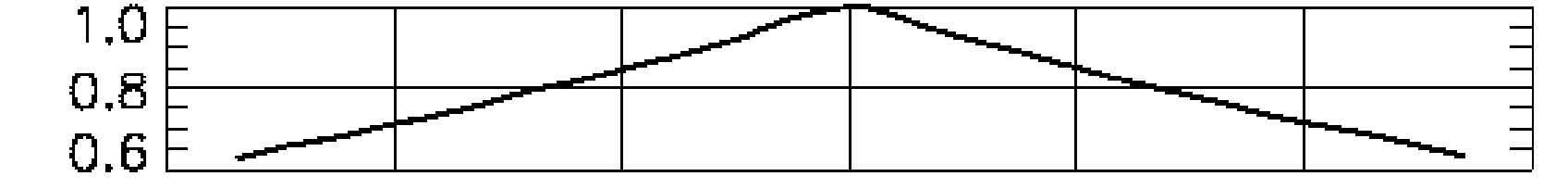}
      \end{flushright}
    \end{minipage}

    \begin{minipage}[c]{0.03\textwidth}
      \begin{flushleft}
	\begin{sideways}{}$\gamma$\end{sideways}
      \end{flushleft}
    \end{minipage}%
    \begin{minipage}[c]{0.45\textwidth} 
      \begin{flushright}
	\includegraphics[width=\textwidth]{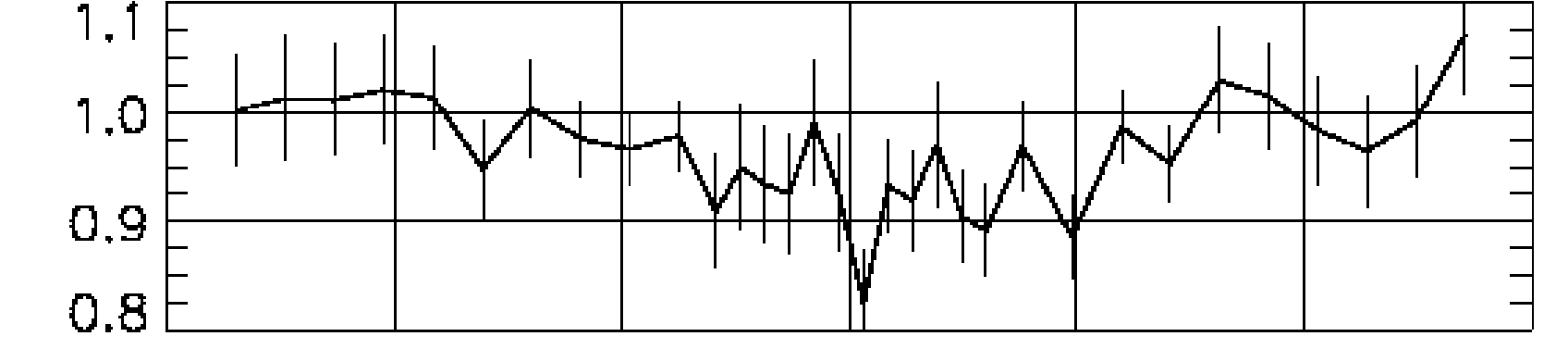}
      \end{flushright}
    \end{minipage}

    \begin{minipage}[c]{0.03\textwidth}
      \begin{flushleft}
	\begin{sideways}{}v (km/s)\end{sideways}
      \end{flushleft}
    \end{minipage}%
    \begin{minipage}[c]{0.45\textwidth} 
      \begin{flushright}
	\includegraphics[width=\textwidth]{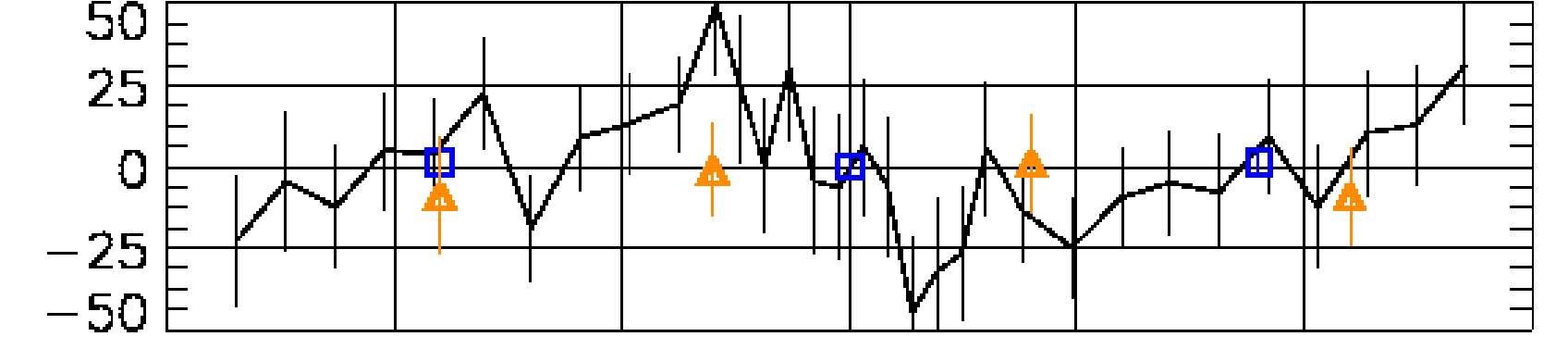}
      \end{flushright}
    \end{minipage}

    \begin{minipage}[c]{0.03\textwidth} 
      \begin{flushleft}
	\begin{sideways}{}$\sigma$ (km/s)\end{sideways}
      \end{flushleft}
    \end{minipage}%
    \begin{minipage}[c]{0.45\textwidth}
      \begin{flushright}
	\includegraphics[width=\textwidth]{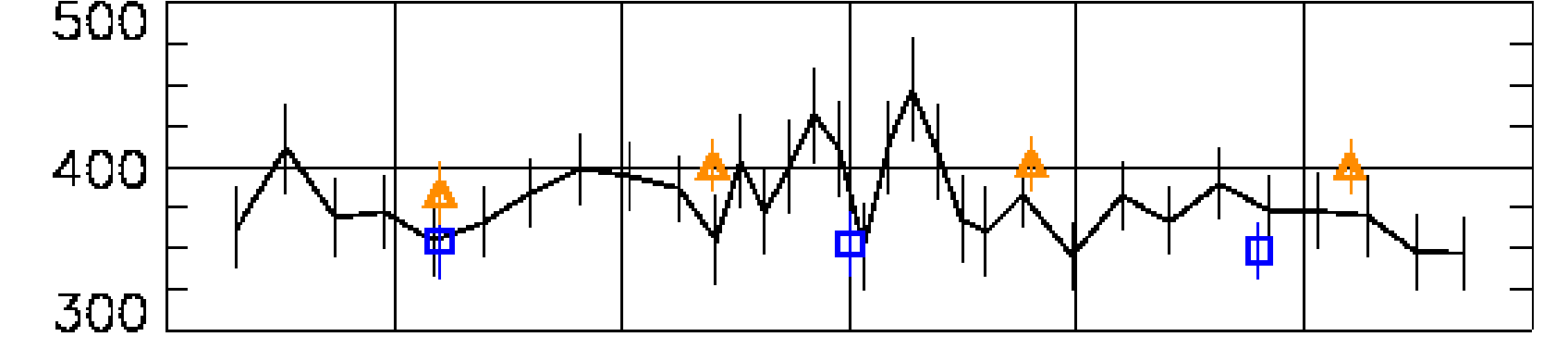}
      \end{flushright}
    \end{minipage}

    \begin{minipage}[c]{0.03\textwidth} 
      \begin{flushleft}
	\begin{sideways}{}$\gamma h_0$\end{sideways}
      \end{flushleft}
    \end{minipage}%
    \begin{minipage}[c]{0.45\textwidth} 
      \begin{flushright}
	\includegraphics[width=\textwidth]{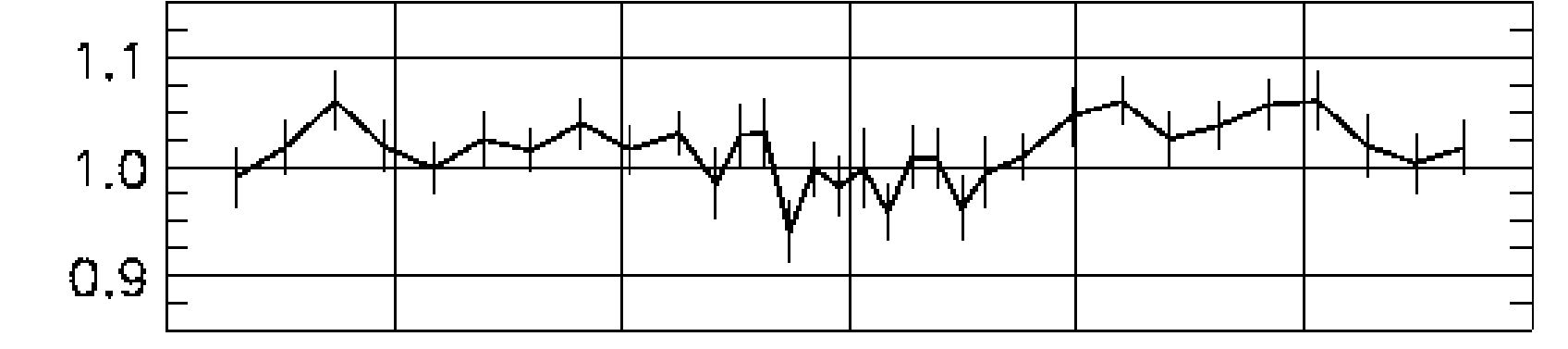}
      \end{flushright}
    \end{minipage}

    \begin{minipage}[c]{0.03\textwidth} 
      \begin{flushleft}
	\begin{sideways}{}$\gamma h_1$\end{sideways}
      \end{flushleft}
    \end{minipage}%
    \begin{minipage}[c]{0.45\textwidth} 
      \begin{flushright}
	\includegraphics[width=\textwidth]{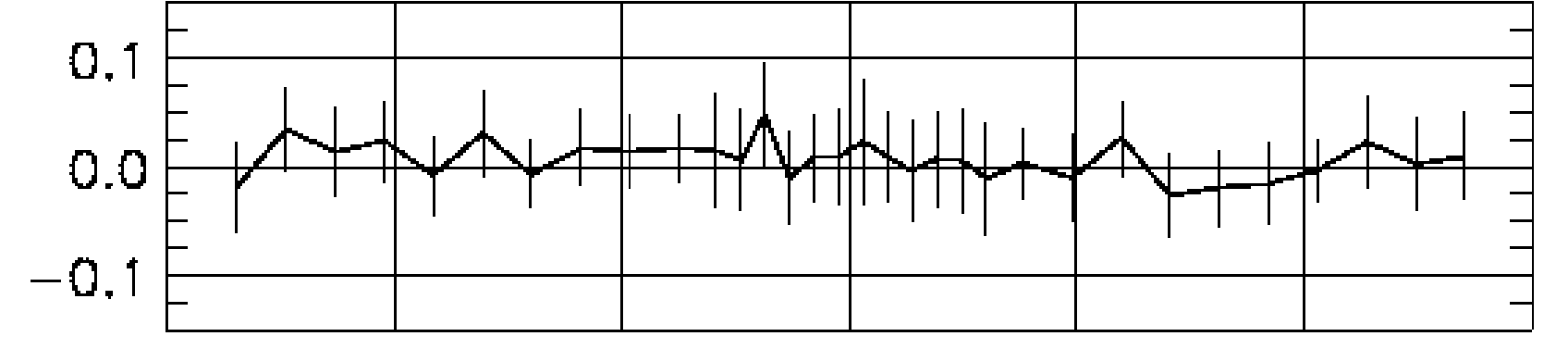}
      \end{flushright}
    \end{minipage}

    \begin{minipage}[c]{0.03\textwidth} 
      \begin{flushleft}
	\begin{sideways}{}$\gamma h_2$\end{sideways}
      \end{flushleft}
    \end{minipage}%
    \begin{minipage}[c]{0.45\textwidth} 
      \begin{flushright}
	\includegraphics[width=\textwidth]{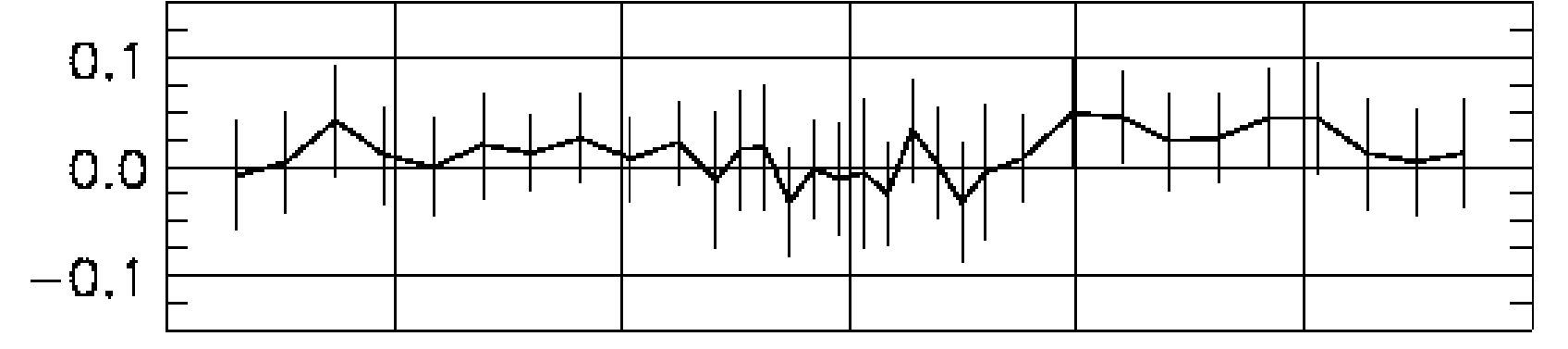}
      \end{flushright}
    \end{minipage}

    \begin{minipage}[c]{0.03\textwidth} 
      \begin{flushleft}
	\begin{sideways}{}$\gamma h_3$\end{sideways}
      \end{flushleft}
    \end{minipage}%
    \begin{minipage}[c]{0.45\textwidth} 
      \begin{flushright}
	\includegraphics[width=\textwidth]{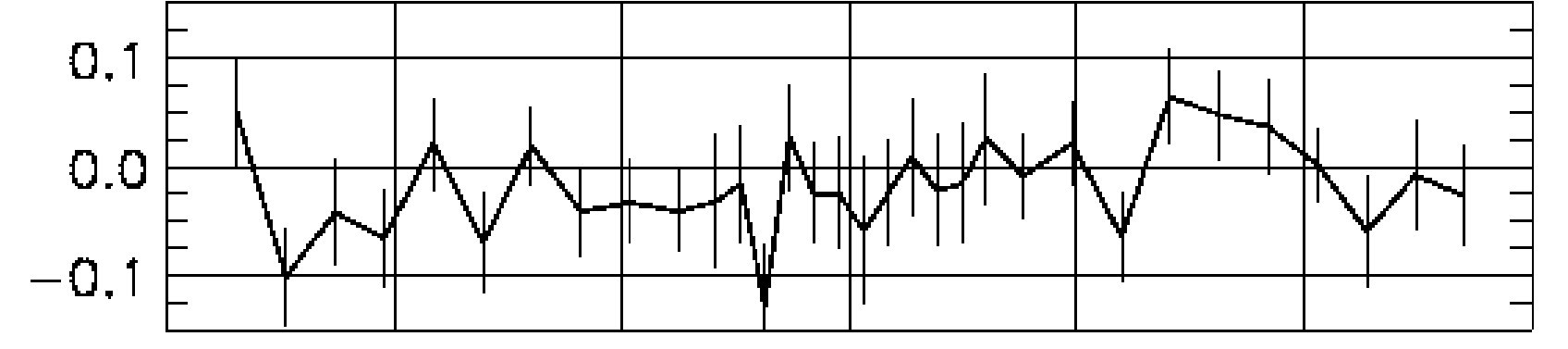}
      \end{flushright}
    \end{minipage}

    \begin{minipage}[c]{0.03\textwidth} 
      \begin{flushleft}
	\begin{sideways}{}$\gamma h_4$\end{sideways}
      \end{flushleft}
    \end{minipage}%
    \begin{minipage}[c]{0.45\textwidth}
      \begin{flushright}
	\includegraphics[width=\textwidth]{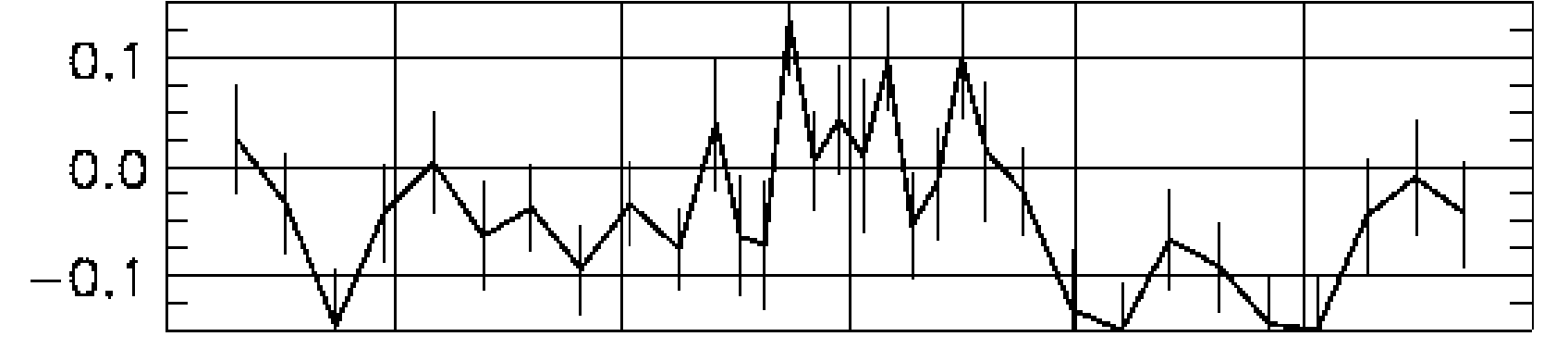}
      \end{flushright}
    \end{minipage}

   \begin{minipage}[c]{0.03\textwidth} 
     \begin{flushleft}
       \begin{sideways}{}W$_{{\rm CO}(2-0)}$\end{sideways}
     \end{flushleft}
    \end{minipage}%
    \begin{minipage}[c]{0.45\textwidth}
      \begin{flushright}
	\includegraphics[width=\textwidth]{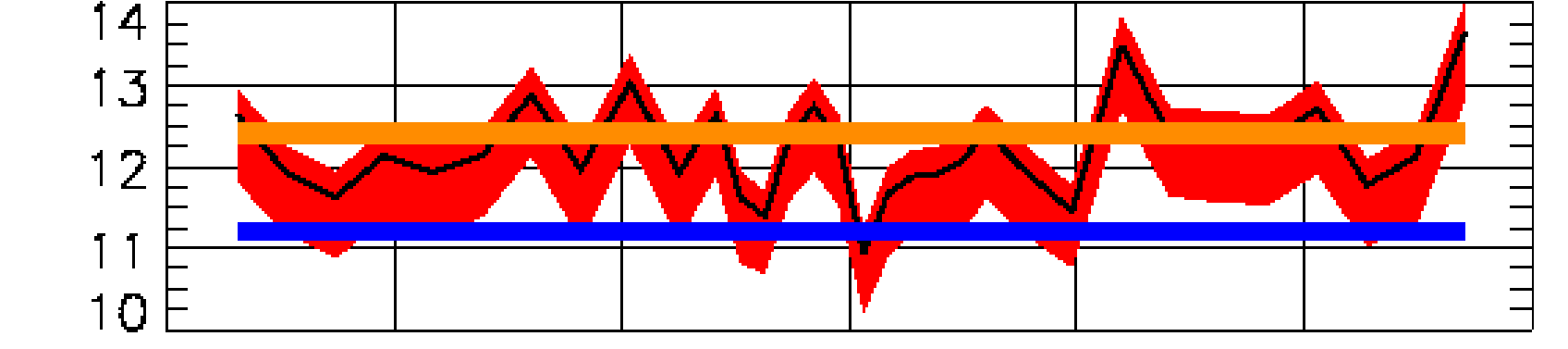}
      \end{flushright}
    \end{minipage}

   \begin{minipage}[c]{0.03\textwidth}
     \begin{flushleft}
       \begin{sideways}{}~~~Templ Frac.\end{sideways}
      \end{flushleft}
    \end{minipage}%
    \begin{minipage}[c]{0.45\textwidth}
      \begin{flushright}
	\includegraphics[width=\textwidth]{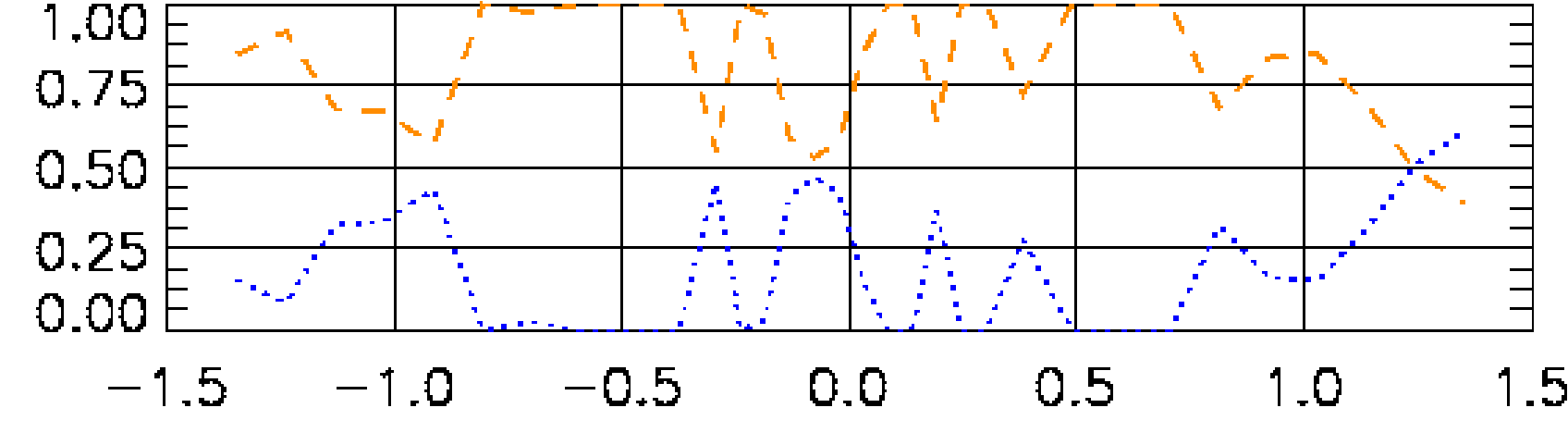}
      \end{flushright}
    \end{minipage}

    \begin{minipage}[c]{0.03\textwidth} 
      \centering{~}
    \end{minipage}%
    \begin{minipage}[c]{0.4\textwidth} 
      \centering{~~~~~~~~Radius from Nucleus (arcsec)}
    \end{minipage}

    \caption{The nuclear kinematics of NGC 1399: The top plot indicates the flux received at each position on the slit, the next 8 plots show the variations in the parameters of (\ref{eq:GHseries}) and the lower two show the template and galaxy EWs and the optimal kinematic template composition as a function of radius from the nucleus. Where applicable, previous data is shown as orange triangles \citep{lon94} and blue squares \citep{graham98}. In the EW plot, the kinematic template EWs are shown in orange (M0III) and blue (K4III); the width represents a 20\kms~error in $v$ and $\sigma_{inst}$. The EW of NGC 1399 is in shown in black (red shows the combined effect of 20\kms~errors in $v$ and $\sigma$). The PA of the slit was $5.06^\circ$ so that the positive radius is northward.}
 
    \label{fig:paramresults}

  \end{figure}
\end{center}


\begin{figure}

~\\
~\\
~\\
\center{See seperate figure}\\
~\\
~\\
~\\

   \caption{The spectra and VPs of the central arcsecond of NGC 1399. The centroid of the light profile along the slit (taken to be the nucleus) is roughly in between two pixels allowing for the VPs to be constructed at approximately equidistant intervals to the left and right of the nucleus, without interpolation. For every point along the slit, two VPs are plotted: black is the Gauss-Hermite VP (\ref{eq:GHseries}) and blue is the non-parametric VP. The velocity axis of the VPs has been corrected to the mean systematic velocity of the system. Accompanying every spectrum is the best-fit broadened template (green) reconstructed from the Gauss-Hermite VP.}
  \label{fig:losvds}
\end{figure}


\section{Results: Imaging and Kinematics}
\label{sec:results}

An 80 second, AO corrected, Ks exposure of NGC1399 is shown in Fig. \ref{fig:1399image}. It has the same PA as the long-slit observations and a slit image (0\farcs172 wide and centred on the AO reference star) has been overlaid. If the reference star is perfectly centred in the slit, there is a maximum position error of 0.3 pixels (1.6pc) on the brightest region of the galaxy from error in the PA. However, the acquisition images (incorporating flat fielding etc. and using the slit images) indicate that the centring of the star along the minor axis of the slit was accurate to only a pixel (5.2pc); the slit was approximately 3 pixels wide (0\farcs172 or 17pc). Hence, the dominant error in aligning the slit on the galaxy centre is from shifts along the minor axis, not from error in the PA. The former is expected to be random about the centre of the slit; the later would give a systematic difference for all observations. The final position accuracy is therefore good: the slit was aligned on the brightest point of the galaxy to an accuracy of around $\pm0\farcs06$ ($\pm6$pc), with no significant systematic offset.

Two globular clusters are also seen in the image data close to the centre of the galaxy. The nearest is 1\farcs15 to the east of the galaxy centre. A Gaussian fit to this globular cluster yields an estimate on the FWHM of the \emph{image} PSF to be 0\farcs078 (7.5pc). This is equal to the \emph{on-source} FWHM predicted by the NAOS preparation software in Sec. \ref{sec:AO}, suggesting that the single Gaussian was probably fit to the diffraction limited component of the globular cluster profile and ignored the seeing limited halo, although the atmospheric conditions were excellent around the time of observation, with the seeing occasionally dropping below 0\farcs4.

The parametrised kinematics of the central few arcseconds of NGC 1399 are shown in Fig. \ref{fig:paramresults}. Where possible, previous data is over plotted \citep{graham98, lon94}. The flux is calculated by summing over all available wavelengths at each point along the slit. The parameters $(\gamma,v,\sigma)$ are chosen to be the best-fit Gaussian parameters and the Gauss-Hermite coefficients $h_n$ are derived using these best-fit Gaussian parameters. The velocity $v$ is given relative to the mean heliocentric velocity of 1467 $\pm$ 4 \kms. Note the error quoted for this velocity is an estimate of the random error only.

Fig. \ref{fig:losvds} compares the non-parametric and parametric VPs for the central arcsecond of the galaxy. Error bars are given for the non-parametric VPs. The galaxy spectra and best-fit broadened template (constructed from the Gauss-Hermite VP) for the central arcsecond are also shown.

\subsection{Highlights}

Many interesting features are present in the above figures. The parametrised kinematics show a decoupled velocity structure across the galaxy centre, a double peaked velocity dispersion across the centre separated by 0\farcs2 or 19pc (with the dip in $\sigma$ half a pixel off the galaxy centre) and significant variations in the $h_i$ with radius. The imaging shows offset asymmetric isophotes at the galaxy centre and the non-parametric VPs (while generally in good agreement with the parameteric VPs) show strong high velocity wings at $r=-0\farcs08$ and lopsided velocity structure at $r=0\farcs14$, which are not detected with the parametrised fit. 

\subsection{Further Details}
\label{sec:fd}
The kinematic features described above persisted when the spectra were reduced using the standard sky subtraction technique for ABBA sequence observations (rather than using an interpolated sky), albeit with increased noise.
Angular distances in Figs. \ref{fig:paramresults} and \ref{fig:losvds} have been shifted such that the brightest region of the galaxy lies at $r=0$\arcsec. This position was determined by scaling to the (outer) centroid of a Nuker profile \citep{Lauer95} fitted to the flux of Fig. \ref{fig:paramresults}.
The photometry is not symmetric at the centre which could bias the calculation, although when the inner 0\farcs5 of the light profile were omitted the change in the centroid was less than 0.1 spectroscopic pixels (5mas or 0.5pc). The centroids of the ellipses fitted to Fig. \ref{fig:1399image} are not constant and fluctuations of up to 0.5 image pixels (13.5 mas or 1.3pc) from the mean are present. The scale on Fig. \ref{fig:1399image} is aligned to the mean x-axis centroid of the ellipses. Registration between the angular scale on the image and that of the kinematics may not be perfect as the zero points were determined independently. However, is likely to be no worse than 0\farcs05.

EW measurements for the two kinematic templates and the galaxy are shown in Fig. \ref{fig:paramresults}. The template EWs (orange for the M0III and blue for the K4III) allow for $\pm20$\kms\ errors in systemic velocity and instrumental dispersion. The corrected EW of the galaxy is shown in black while the effect of $\pm20$\kms\ errors in velocity and dispersion are shown in red. The asymmetry in the error is associated with the error in velocity: large positive or negative changes in velocity move the CO (2-0) feature out of the predefined wavelength limits \cite{omo93}, thus reducing the EW. The kinematics $(v,\sigma)$ used to correct the galaxy EW are those shown in Fig. \ref{fig:paramresults}. Systematic differences in $(v, \sigma)$ from using different templates in kinematic extraction are less than 20\kms~in each case. 
These EW measurements indicate that our kinematic templates are well matched to the luminosity weighted stellar population of the galaxy. With reference to \citet{omo93}, one can see that the corrected EW measurements of the two templates (11.3 and 12.5 for the K4III and M0III, respectively) are typical of their spectral classifications. The galaxy EW (average value of 12.1) is closer to that of the M0III template; this is also reflected by the favouring of the M0III template in the kinematic extraction (Fig. \ref{fig:paramresults}). However, one should be apprehensive about the variations of the galaxy's EW with radius: the dispersion correction assumes a Gaussian VP which we have seen is not always the case. Overall, the analysis of EWs indicates that the templates are well matched to the galaxy population, so we should not expect problems associated with template mismatch to be present in the kinematics. Additionally, previous (seeing limited) data on the central velocity dispersion of NGC 1399 agrees with our data (Fig. \ref{fig:paramresults}), given the different resolutions. \citet{saglia00} report slightly negative $h_4$ towards the galaxy centre which we confirm although our rise in $h_4$ in the very central 0\farcs5 is beyond the (seeing limited) spatial resolution of their data.

To maintain a high SNR at larger radii ($ \mid{r}\mid{} > 0\farcs3$), the spectra were binned into pairs prior to kinematic extraction. However, after binning, $\chi_{\rm S}^2$ (\ref{eq:chisqspec}) rose, on average, from $\sim$160 to $\sim$230 (an approximate change of $\sqrt2$) for a fit to 174 data points. This indicates that systematic errors are starting to become comparable to the random errors. There are many possible causes of systematic errors. 
The optimal template, convolved with the best VP will not reproduce all the
features in the galaxy spectrum; although the galaxy and template spectra   
have comparable CO (2-0) EWs, there may be absorption or even
emission features in the galaxy spectra that are not obvious and are not
accounted for in eq.~(\ref{eq:chisqspec}), but still contribute to the
differences between the galaxy and templates and to the systematic
`noise'.
Telluric correction is only estimated to be accurate to 1\%, increasing to 2\%-3\% at regions where the telluric absorption shows sharp prominent features. In addition, the stars used to obtain a telluric spectra are unlikely to have featureless spectra at high SNRs. The SNR of the data binned at 0\farcs3 over the wavelength range used to extract kinematics falls from 110 (2.25\mum) to 30 (2.45\mum) per pixel due to the thermal background (the SNR at the unbinned galaxy centre falls from 90 to 20). This SNR (and corresponding error estimates) quantify random error alone and do not account for systematics introduced by telluric correction. Binning will reduce the random noise but not the systematic noise. Thus as $\chi_{\rm S}^2$ (\ref{eq:chisqspec}) is weighted by the random noise estimates, when the random error in each spectrum is reduced, it will increase so long as the systematic noise persists.

\section{Discussion}
\label{sec:datadiscus}

The interpretation of the results is discussed below in appropriate sections. 

\subsection{Decoupled Kinematics}
\label{sec:dis:dk}
There is a strong rotation gradient in $v$ within a radius of $0\farcs5$ (48pc) which is clearly decoupled from the kinematics at larger radii. The  magnitude of the central rotation (taking the difference of the maximum velocity in each direction) is $\sim$70\kms. The absence of high resolution data perpendicular to our long slit position prevents us from concluding if the system is truly counter rotating, or just decoupled. Previous publications considered the possibility of a decoupled system \citep{don95,saglia00}, but the poorer spatial resolution of the data prevented a reliable detection. Kinematically decoupled cores are often (but not exclusively) found in spherical, high dispersion systems with core-like photometry \citep{sauronIII}.

\subsection{The Central Dispersion Profile}

The drop in $\sigma$ half a pixel off the galaxy centre is accompanied by less convincing dips in $\gamma$ and the CO (2-0) EW. These features are close to the limit of the spatial resolution and thus may be unresolved. It is well known that random errors in $\gamma$ and $\sigma$ are statistically correlated \citep{EfEC1980} and these features are only clearly seen when continuum correction is incorporated into the minimisation of $\chi_{\rm S}^2$ (\ref{eq:chisqspec}); if continuum correction is left out of the minimisation (i.e. calculated once from the continuum shortward of the CO (2-0) feature), $\gamma$ shows no obvious change (and is noisier in general) and the dispersion becomes flat at the centre.

The decrease in $\gamma$ and $\sigma$ could be correlated effects from fitting constrained parametrised VPs to more complicated profiles; $\gamma$ is expected to fall at the centre \citep{vdM94}. However, the non-parametric VP at $r=0\farcs03$ matches the parametric form well. Alternatively, the dip in $\sigma$ could reflect a genuine fall in the dispersion of the stars at the centre of the galaxy, but this is the exact opposite to what one would expect when approaching the BH.

An alternative explanation is the presence of young stellar population at the centre, although $\gamma$ and $\sigma$ may not necessarily decrease in this case: a young stellar population may include younger MK types with reduced CO (2-0) EWs, but it would also have an increased fraction of supergiants, which have a higher CO (2-0) EW than giants for the same MK spectral type \citep{omo93}. In addition, one would expect an correlated change in the optimal template used to extract the kinematics, which is not the case here. One would expect a different stellar population to produce a colour gradient but unfortunately there is insufficient homogeneous data at the required resolution to confirm this.

Central dips in $\sigma$ have been seen in other core galaxies such as M87 and NGC 4649 \citep{vdM94,pink2003}, which both have nuclear activity. If NGC 1399 contained a central \emph{nucleus} of light from a non-thermal, non-stellar object \citep{vdM94}, one would expect the CO bands to show a lower $\gamma$ relative to the continuum level which may introduce correlated dips in $\gamma$ and $\sigma$. New photometry by \citet{Lauer05} shows a slight excess of light at the galaxy centre after subtraction of the best-fit Nuker profile and therefore suggests the presence of a very faint nucleus. However, the authors emphasise the possible dangers and unknown systematics of extrapolating the Nuker fit into the central regions.

\subsection{Offset Photometry and Peculiar VPs}

The Ks band image of NGC 1399 shows a departure from spherical symmetry in the central 0\farcs5 with an elongation of the surface brightness towards the east-south-east, approximately 0\farcs2 (19pc) in length. The image is a relatively short exposure and the SNR is low. However, the elongation of the central isophotes covers more than 16 pixels; the probability of finding 16 neighbouring pixels above the mean, given the noise statistics of the image, is negligible so we can rule out the possibility of random error causing such an artifact. Obvious systematic errors which may cause elongation of the central isophotes would be either an error in frame alignment, or an unusual PSF from the AO correction. However, both these effects can be dismissed: the globular cluster found 1\farcs15 from the galaxy centre is circular and well fit by a 2D Gaussian of equal width in each dimension; no significant residual is seen after subtraction. 
We note that HST H-band images exists for NGC1399, but
owing to a poorer spatial resolution (0\farcs13) and PSF sampling, these structures are
not observed. 
The non-spherical isophotes at the centre of NGC 1399 are a genuine feature of the galaxy. There appear to be peculiar kinematic features associated with the photometric anomaly: at $r=-0\farcs08$, the non-parametric VP contains substantial high velocity wings and at $r=0\farcs14$, the non-parametric VP is asymmetric with an excess of receding velocity structure. We discuss possible explanations for these features below.

Isophote twists are common in core-regions of core galaxies \citep{Lauer05}. However, such twists are generally smooth and seen on large scales, which is not the case here. The elongation could be a projection effect from the obscuration by a non-uniform distribution of dust. However, no obvious dust signature can be seen in either this Ks band image or archival HST V-band images. The density distribution of the dust would need to vary rapidly to cause such a sudden effect in the NIR on such small spatial scales (0\farcs3 or 29pc). However, neither of these explanations would give rise to the associated kinematic features.

We have already seen that globular clusters (GCs) have been resolved near the centre of NGC 1399. Although the elongation of the nucleus does not appear to be separate to the central maximum, it is plausible that a GC happens to lie in the line of sight between ourselves and the galaxy centre and is unresolved from the central photometric maximum of the galaxy. There would be a kinematic effect from such a chance alignment: one would expect a `spike' in the VP at the systematic velocity of the GC caused by the low dispersion velocity structure of the globular cluster. However, we see high velocity wings in the VP at $r=-0\farcs08$ and a lopsided structure at $r=0\farcs14$ which is not the same.

The elongation of the central isophotes is more likely to be an offset centre as seen in a handful of other \emph{core} galaxies \citep{Lauer05}. In fact, \citet{Lauer05} argue that the offset centres detected in new WFPC2 photometry are all eccentric disks, analogous to that of M31 \citep{tremaine95}. Certainly, this would produce a strong kinematic signature in the form of high velocity wings if our slit bisected the eccentric disk. Our slit does not completely bisect the peculiar photometry and nor does it coincide with the PA of the eccentric isophotes. However, the alignment of the AO reference star in the slit varied by around 1 pixel and this translates to a potential \emph{wandering} of the slit, perpendicular to its length, by two image pixels. Thus, there is likely to be considerable contribution from the elongated photometry in the spectroscopic data. 

Keeping in mind the problems with registration between the image and kinematics, an eccentric disc at the centre of NGC 1399 would create high velocity wings at $r=-0\farcs08$, where the slit is closest to bisecting the anomaly. If the non-parameteric VP at this point is to be trusted, the velocity dispersion of just the wings (ignoring the central bulk) is in excess of 1000\kms. Furthermore, if the stars in the eccentric disc rotate in the same direction, one would expect a lopsided VP at the pericentre of the ellipse if the disk was viewed near edge on. We do see a strong excess or `hump' of structure receding away from the observer at $r=0\farcs14$ which is possibly the closest pixel to the pericentre of the elliptical photometry. Although, \citet{Lauer05} report no offset photometry in NGC 1399 with new WFPC2 photometry, the isophote ellipticity is seen to jump from approximately 0 to 0.2 between $0\farcs09 < r < 0\farcs1$.

We note that this eccentric disk hypothesis cannot explain the decoupled kinematics discussed in Sec. \ref{sec:dis:dk}. A coherent disk can only survive well within the SoI of the BH, which we estimate to be $\sim0\farcs3$.


\section{Dynamical modelling}
\label{sec:modelling}

To obtain preliminary constraints on the mass of any BH in NGC~1399,
we ignore the mild rotation gradient and elongated central isophotes
and fit spherical dynamical models to our central kinematics combined
with \citet{graham98}'s velocity dispersion profile.  The latter
extends to 70 arcsec and provides important constraints on the
galaxy's mass-to-light ratio.

We assume that mass follows light, except at the galaxy centre where
there can be a BH.  The mass density distribution is then
\begin{equation}
\label{eq:densityprofile}
\rho(r) = M_\bullet\delta(r) + \Upsilon j(r).
\end{equation}
Our goal is to find the range of BH masses $M_\bullet$ and stellar
mass-to-light ratios~$\Upsilon$ that are consistent with our
kinematics.  We take $j(r)$ from the models of \citet{mag98}, which
was obtained by deprojecting a composite surface brightness profile
constructed by combining HST and ground-based photometry.  Having this
$j(r)$ it is straightforward to calculate the gravitational
potential~$\psi(r)$ corresponding to~(\ref{eq:densityprofile}) for any
choice of $M_\bullet$ and $\Upsilon$.  Throughout this paper all
mass-to-light ratios are in the $V$ band.

Our modelling procedure is a straightforward adaptation of the
extended Schwarzschild method described by \citet{rix97} and
\citet{cretton99}:
\begin{enumerate}
\item choose trial values for $M_\bullet$ and~$\Upsilon$ and calculate
  the corresponding potential~$\psi(r)$;
\item follow a representative sample of orbits in this $\psi(r)$;
\item find the weighted combination of orbits that minimises the
  $\chi^2$ of the fit between the model and the observations,
  subject to the constraint that each orbit carries non-negative
  weight;
\item assign the likelihood $\exp(-{1\over2}\chi^2)$ to the potential~$\psi$.
\end{enumerate}
We do not impose any regularization in the third step.  By
considering a range of plausible trial potentials and comparing their
relative likelihoods, we obtain constraints on both $M_\bullet$
and~$\Upsilon$.

\subsection{Orbit distribution}
\label{sec:modelgrid}
Apart from the parameters $(M_\bullet,\Upsilon)$ describing the
potential, the other unknown in our models is the distribution
function (DF) $f(\b x,\b v)$, defined such that $f(\b x,\b v)\,\d^3\b
x\d^3\b v$ is the luminosity of stars in some small volume
$\d^3\b x\d^3\b v$ of phase space.  Jeans' theorem tells us that if
our models are to be in equilibrium their DFs can depend on $(\b x,\b
v)$ only through the integrals of motion $\E$ and $\b J$, the binding
energy and angular momentum per unit mass.  We make the stronger
assumption that the local stellar velocity distribution is symmetric
about the $v_r$ axis, so that the DF depends on $\b J$ only through
its magnitude.  We then discretize the DF as a sum of delta functions,
\begin{equation}
f(\E,J^2) = \sum_{i=1}^{n_\E}\sum_{j=1}^{n_J} f_{ij}
\delta(\E-\E_i)\delta(J^2-J^2_{ij}),
\label{eq:DFdoublesum}
\end{equation}
on a regular $n_\E\times n_J$ grid in phase space.  The points $\E_i$
are chosen through $\E_i=\psi(r_i)$ with the $r_i$ spaced
logarithmically between 1 pc and 100 kpc.  This ensures that orbit
apocentres are approximately uniformly distributed among each decade
in radius. For each~$\E_i$ there are $n_J$ values of angular momentum,
with $J^2_{ij}$ running linearly between $0$ and $J^2_{\rm c}(\E_i)$,
the angular momentum of a circular orbit of energy~$\E_i$.  To avoid a
rash of indices we also write the double sum~(\ref{eq:DFdoublesum}) as
a single sum over $n\equiv n_\E\times n_J$ points:
\begin{equation}
f(\E,J^2) = \sum_{k=1}^n f_k \delta(\E-\E_k)\delta(J^2-J^2_k).
\label{eq:DFdiscretization}
\end{equation}

\subsection{Observables}
\label{sec:modelproj}
Having a trial potential $\psi(r)$ and a set of DF components
$(\E_k,J^2_k)$, we calculate the unnormalised, psf-convolved VP
histogram of each component at the projected radius~$R_i$ of each of
our $n_{\rm AO}=31$ kinematical data points:
\begin{eqnarray}
L_{ij}^{(k)}&& = \int_{v_j}^{v_{j+1}}\,\d v_z\int\d x\d
y\,\psf(R_i-x,-y)\cr
&&\qquad\times\int\d v_x\d v_y \delta(\E-\E_k)\delta(J^2-J^2_k).
\label{eq:vpcoeff}
\end{eqnarray}
Here we use a rectangular co-ordinate system $(x,y,z)$ with origin $O$
at the galaxy centre and $Oz$-axis parallel to lines of sight.  For
our standard models each histogram has $n_v=24$ bins of width
$v_j-v_{j-1}=50$\kms, with innermost bin edge at $v_1=0$\kms, the
galaxy's systemic velocity, and outermost edge at $v_{25}=1200$\kms. The function $\psf(\Delta x,\Delta y)$ is the two-dimensional off-source point-spread function of Sec.~\ref{sec:AO} and Table \ref{tab:psfs}.  Armed with the
$L_{ij}^{k}$, the unnormalised VP histogram of a model with DF
$(f_1,\ldots,f_n)$ is simply
\begin{equation}
L(R_i;v_j,v_{j+1}) = L_{ij} = \sum_{k=1}^n f_k L_{ij}^{(k)}.
\label{eq:vpsum}
\end{equation}
The normalisation constant is the psf-convolved surface brightness,
\begin{equation}
I(R_i) = I_i = \sum_{k=1}^n f_k I_i^{(k)},
\label{eq:sbsum}
\end{equation}
where, by analogy with~(\ref{eq:vpcoeff}), 
\begin{equation}
I_i^{(k)} = \int\d x\d
y\,\psf(R_i-x,-y)\int\d^3\b v \delta(\E-\E_k)\delta(J^2-J^2_k).
\label{eq:sbcoeff}
\end{equation}
We explain how we evaluate the multiple integrals (\ref{eq:vpcoeff})
and~(\ref{eq:sbcoeff}) in \citet{mag05}.  The spherical symmetry of
our models means that we can afford to calculate these projection
coefficients for each of our 31 observed radii directly.  More
sophisticated axisymmetric models (e.g., \citep{cretton99,geb03})
usually resort to introducing subgrids to store intermediate
quantities in this calculation.

Our treatment of Graham et al's velocity dispersion profile is
similar.  We assume that each of their data points measures the second
moment of the VP convolved with a Gaussian PSF with FWHM 2 arcsec.
The (unnormalised) second moment of each DF component is given
by~(\ref{eq:sbcoeff}) with an extra factor of $v_z^2$ inside the
innermost integral.

\subsection{Fitting models to observations}
\label{sec:modelfit}

\subsubsection{Gauss--Hermite coefficients}
\label{sec:modelfitgh}
Calculating the Gauss--Hermite coefficients of our models is simple:
the (unnormalised) contribution of the $k^{\rm th}$ DF component to
the $i^{\rm th}$ VP is given by $L_{ij}^{(k)}$, with $j=1,\ldots,n_v$,
from which equation~(\ref{eq:GHcoeff}) allows us to calculate this
component's contribution $h_{ij}^{(k)}$ to $\{h_j\}$.  However, as we
have explained in section~\ref{sec:ghvps}, the observed Gauss--Hermite
coefficients are not independent.  It takes only a little effort to
include the effects of the covariances among the $\{h_j\}$ into the
modelling.

Let us define the column vector $\b h\equiv(h_0,\ldots,h_N)^T$ and let
$\hat\b h$ be the vector of coefficients that minimise
the~$\chi^2_{\rm s}$ of equation~(\ref{eq:chisqspec}).  For fixed
normalisation~$k$ and continuum parameters~$c_i$, this $\chi^2_{\rm
  s}$ is a quadratic form in the $h_i$:
\begin{eqnarray}
  \chi^2(\b h) &\simeq& \chi^2_{\rm min} + {1\over2}(\b h-\hat\b h)^T
  \cdot M \cdot (\b h-\hat\b h)\cr
  & = & \chi^2_{\rm min} +
  {1\over2}\sum_{i=0}^N\lambda_i [\b e_i\cdot(\b h-\hat\b h)]^2,
\label{eq:chisqghquad}
\end{eqnarray}
where the Hessian $M_{ij}\equiv \p^2\chi^2_{\rm s}/\p h_i\p h_j$ has
eigenvalues~$\lambda_0,\ldots,\lambda_N$ with corresponding
eigenvectors~$\b e_0,\ldots \b e_N$.  Therefore, the new parameters
\begin{equation}
h'_j\equiv \b e_j\cdot\b h 
\label{eq:sGH}
\end{equation}
have independent errors $\Delta_j\equiv\sqrt{2/\lambda_j}$.

Summing the results from~(\ref{eq:chisqghquad}) for each of our measured
VPs, 
\begin{equation}
\chi^2_{\rm GH} = \sum_{i=1}^{n_{\rm AO}}\sum_{j=0}^N
\left(\gamma_i\hat h'_{ij}-{1\over\hat I_i}\gamma_i\sum_{k=1}^nf_kh'^{(k)}_{ij}
  \over \Delta_{ij}\right)^2,
\label{eq:chisqmodgh}
\end{equation}
where we have dropped the $\chi^2_{\rm min}$ terms and have introduced
the $h'^{(k)}_{ij}$, which are related to the $h_{ij}^{(k)}$
by~(\ref{eq:sGH}).  The model's predicted VPs in~(\ref{eq:chisqmodgh})
are normalised by the ``observed'' local surface brightnesses~$\hat
I_i$.  The latter are difficult to extract from spectroscopic
observations.  Instead, we obtain them by convolving the photometric
profile used to obtain our~$j(r)$ profile with the effective
spectroscopic PSF.  Of course, a reasonable model must also fit
this~$I(R)$, so we add
\begin{equation}
\chi^2_I=\sum_{i=1}^{n_{\rm AO}}\left(
  \hat I_i-\sum_{k=1}^nf_kI_{i}^{(k)}\over\Delta I_i
\right)^2
\end{equation}
to~(\ref{eq:chisqmodgh}), somewhat arbitrarily assigning errors
$\Delta I_i = 10^{-3}\hat I_i$.  Finally, we add one more term,
$\chi^2_G$, to measure how well our models fit both the unnormalised
second moments from Graham et al.'s data and the corresponding
psf-convolved surface brightnesses, again assuming fractional errors
of~$10^{-3}$ in the latter.  Adding all these measurements together,
the final $\chi^2$ of our model,
\begin{equation}
\label{eq:chisqmodall}
\chi^2_{\rm m}[\psi,f_k]=\chi^2_{\rm AO}+\chi^2_{I}+\chi^2_{\rm G},
\end{equation}
which, for fixed potential~$\psi$, depends quadratically on the orbit
weights~$f_k$.  We use a standard non-negative least-squares algorithm
\citep{LH1974} to find the non-negative set of~$f_k$ that
minimise it.  Unlike most other orbit-superposition methods, we do not
include the luminosity density $j(r)$ in this fit explicitly, but use
the surface brightness profile $I(R)$ instead.  Our reason for this is
that, unlike $j(r)$, $I(R)$ is (in principle) directly measurable and
can therefore be assigned meaningful error bars.  We do not expect real
galaxies to have perfectly constant mass-to-light ratios and, in the
absence of anything better, use $j(r)$ merely to estimate the stellar
contribution to the overall potential.

\subsubsection{VP histograms}
\label{sec:modelfithist}

\begin{figure}
\includegraphics[width=0.9\hsize,angle=270]{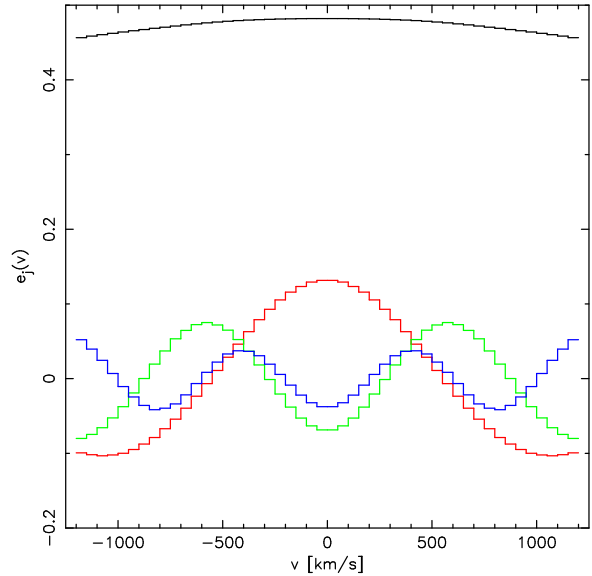}
\caption{Plot of the first four eigenVPs of one of our spectra
  obtained by diagonalizing $\chi_{\rm s}$ of
  eq.~(\ref{eq:chisqspec}).  Any (symmetric) VP can be expressed as a
  weighted sum of eigenVPs (eq.~\ref{eq:eigenvp}), in which case the
  errors in the weights are independent.  We have divided each of the
  VPs plotted here by its corresponding eigenerror so that the scale
  of each gives a direct indication of the uncertainty in its weight.}
\label{fig:eigenvps}
\end{figure}

Unlike the Gauss--Hermite coefficients above, there is no simple way
to compare the VP histograms found in~\S\ref{sec:histvps} against
models: in addition to the unavoidable correlations among the $L_i$,
there are complicated biases introduced by the penalty
function~(\ref{eq:penaltyfn}).  So, we simply refit the VP histograms
using the same $n_v$ velocity bins for which we calculate the models'
VPs in \S\ref{sec:modelproj}, reflecting the VPs about $v=0$ in order
to make them symmetric.  The procedure is as follows:
\begin{enumerate}
\item First find the best-fit template fraction, continuum level and
  normalisation: find the $f$, $c_i$ and smoothed $L_i$ that minimise
  the penalised $\chi^2_{\rm p}=\chi^2_{\rm s}+P[L_i]$ for $k=1$,
  where $\chi^2_{\rm s}$ and the penalty function~$P$ are given by
  equations (\ref{eq:chisqspec}) and~(\ref{eq:penaltyfn})
  respectively.  The best-fit normalisation factor is then
  $k^{-1}=\sum_i (v_{i+1}-v_{i})L_i$.
\item Holding $k$, the $c_i$ and $f$ fixed, find the {\em formal}
  best-fit VP histogram $(\hat L_1,\ldots,\hat L_{n_v})$ to the
  unpenalised~$\chi^2_{\rm s}$~(\ref{eq:chisqspec}).
\end{enumerate}
The best-fit VP~$\hat\b L$ is in general unphysical with many $\hat
L_i<0$, but we use it only because it locates the minimum of the
$\chi^2_{\rm s}$ quadratic form.  Let us write our formal best-fit VP
histogram as the column vector $\hat\b L\equiv(\hat L_1,\ldots,\hat
L_{n_v})^T$ and consider another~$\b L\equiv(L_1,\ldots,L_{n_v})^T$.
Once $k$, $f$ and the $c_i$ have been fixed,
equation~(\ref{eq:chisqspec}) becomes
\begin{equation}
\chi^2_{\rm s}(\b L) = \chi^2_{\rm min} + 
\sum_{i=1}^{n_v} \left(\b e_i\cdot\b L - \b e_i\cdot\hat\b L\over\Delta_i\right)^2,
\label{eq:chisqhistvp}
\end{equation}
where $\b e_i$ are the eigenvectors of the Hessian $\p^2\chi^2_{\rm
  s}/\p L_i\p L_j$ and the eigenerrors~$\Delta_i$ are related to the
eigenvalues~$\lambda_i$ through $\Delta_i\equiv\sqrt{2/\lambda_i}$.
So by taking the projections 
\begin{equation}
L'_i\equiv \b e_j\cdot\b L,
\label{eq:eigenvp}
\end{equation}
of~$\b L$ along the full set of ``eigenVPs''~$\b e_j$, one test
directly how well it reproduces the observed galaxy spectrum.  We note
that our method is essentially a restatement of the work of
\citet{DD98} using the language of eigenVPs.
Figure~\ref{fig:eigenvps} plots the first few eigenVPs of one of our
spectra.  Unlike the terms in a Gauss--Hermite
expansion~(\ref{eq:GHseries}), they do not taper off rapidly at high
velocities.

Summing the results of~(\ref{eq:chisqhistvp}) for each VP and
neglecting the $\chi^2_{\rm min}$ terms, a model with DF
$(f_1,\ldots,f_n)$ has
\begin{equation}
\chi^2_{\rm H} =  \sum_{i=1}^{n_{\rm AO}}\sum_{j=1}^{n_v}
\left(\hat L'_{ij}-{1\over\hat I_i}\sum_{k=1}^nf_kL'^{(k)}_{ij}
  \over \Delta_{ij}\right)^2,
\label{eq:chisqmodhist}
\end{equation}
where $L'^{(k)}_{ij}$ is obtained from $L^{(k)}_{ij}$ through~(\ref{eq:eigenvp}).
Apart from replacing $\chi^2_{\rm GH}$ in
equation~(\ref{eq:chisqmodall}) by~(\ref{eq:chisqmodhist}), our
procedure for fitting models to VP histograms is identical to that for
Gauss--Hermite coefficients.

\subsection{Results}
\label{sec:modelresults}

\begin{figure*}
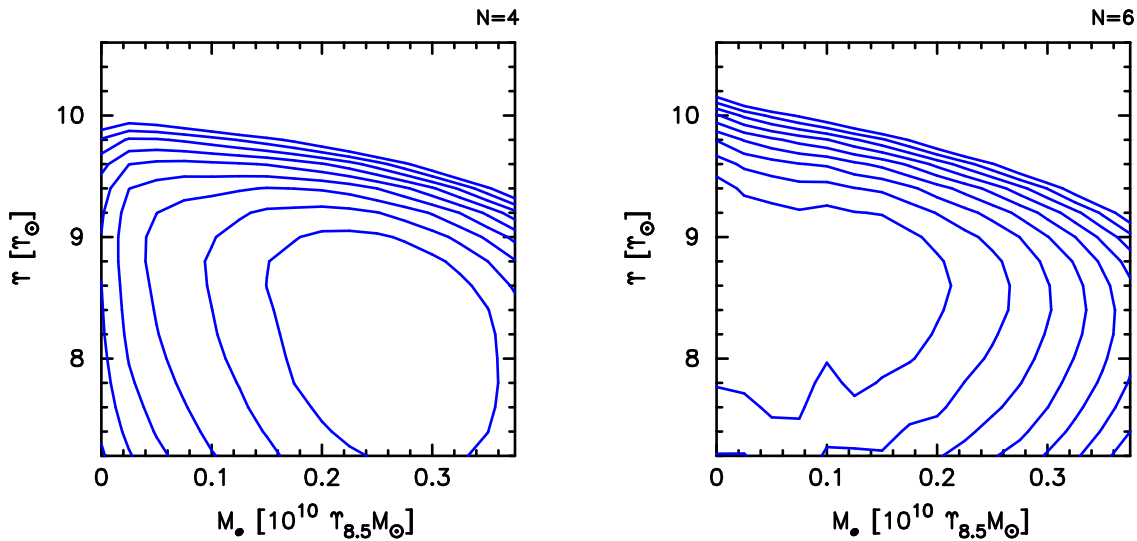

\includegraphics[width=0.4\hsize,angle=270]{fig7a.ps}\hfil
\includegraphics[width=0.4\hsize,angle=270]{fig7b.ps}
\caption{$\chi^2_{\rm m}(M_\bullet,\Upsilon)$ contours
  (eq.~\ref{eq:chisqmodall}) obtained by fitting dynamical models to
  Gauss--Hermite parametrisations (Sec.~\ref{sec:ghvps}) of our
  spectra and to the velocity dispersion profile measured by
  \citet{graham98}.  Successive contour levels have $\Delta\chi^2_{\rm
    m}=1$.  The models have $200\times40$ DF components
  (\ref{eq:DFdoublesum}) and fit $(h_0,h_2,\ldots,h_N)$ extracted
  from spectra under the assumption that $h_i=0$ for $i>N$.  The left
  panel shows results for $N=4$, the right for $N=6$.  The $x$-axis is
  both cases is the BH mass scaled by
  $\Upsilon_{8.5}\equiv\Upsilon/8.5\Upsilon_\odot$. Mass-to-light
  ratios~$\Upsilon$ are $V$ band.}
\label{fig:chisqpotgh}
\end{figure*}

\begin{figure*}
\centerline{
  \includegraphics[width=0.4\hsize,angle=270]{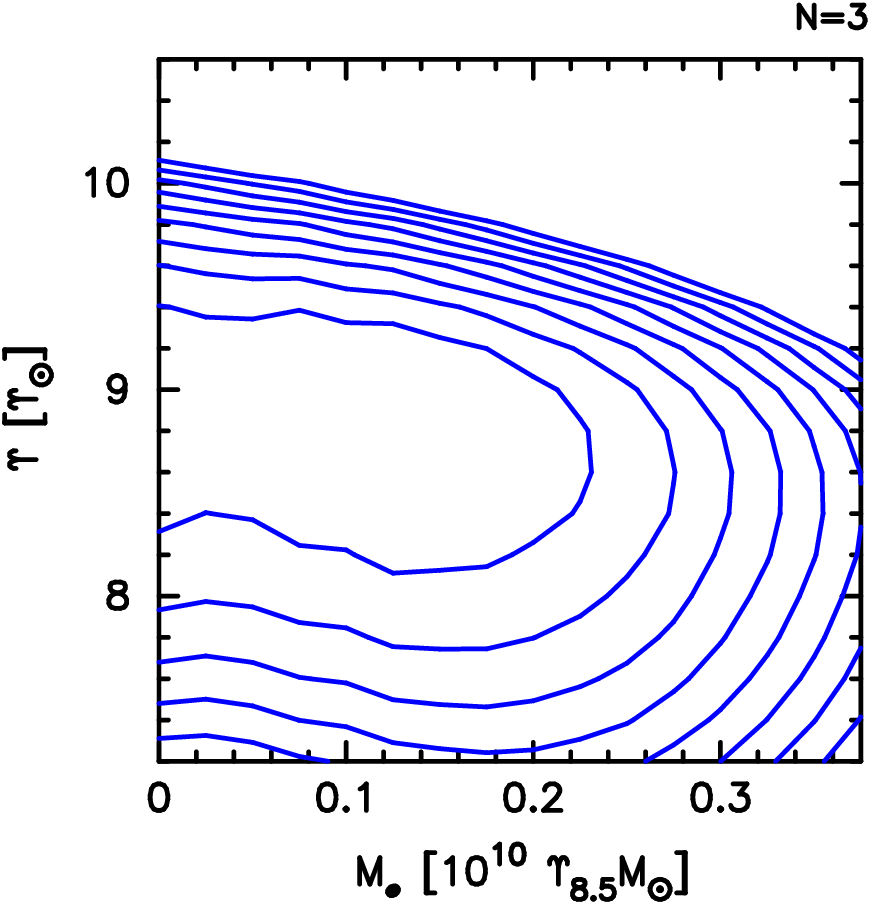}
  \hfil
  \includegraphics[width=0.4\hsize,angle=270]{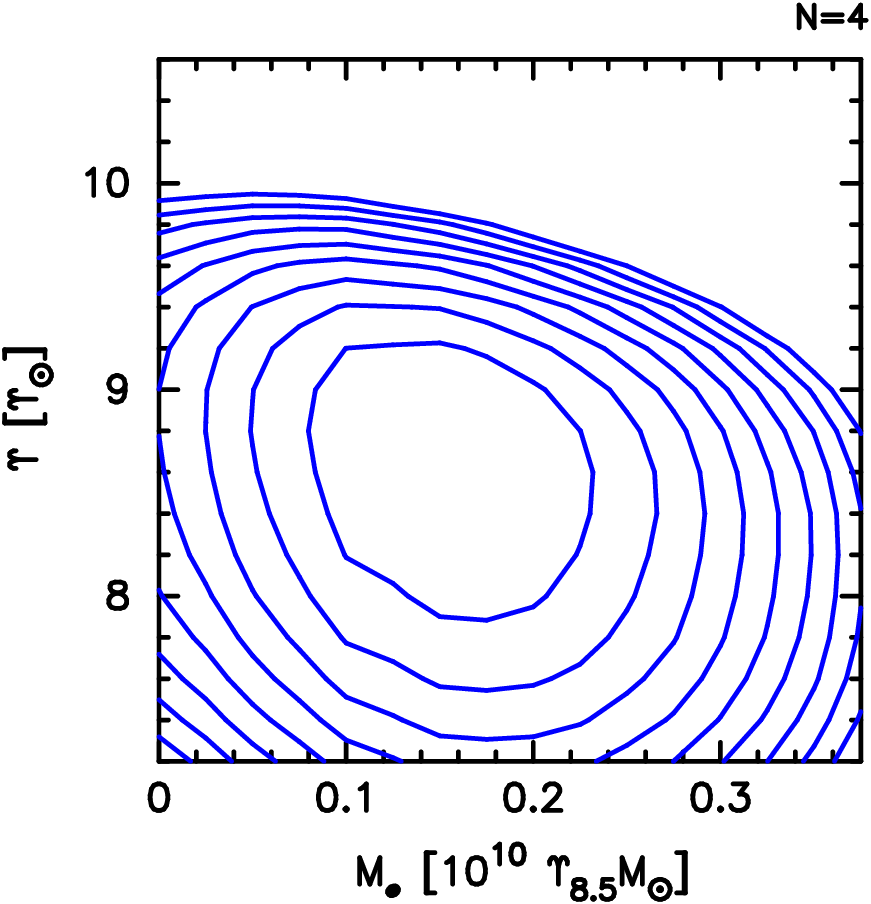}}
\centerline{
  \includegraphics[width=0.4\hsize,angle=270]{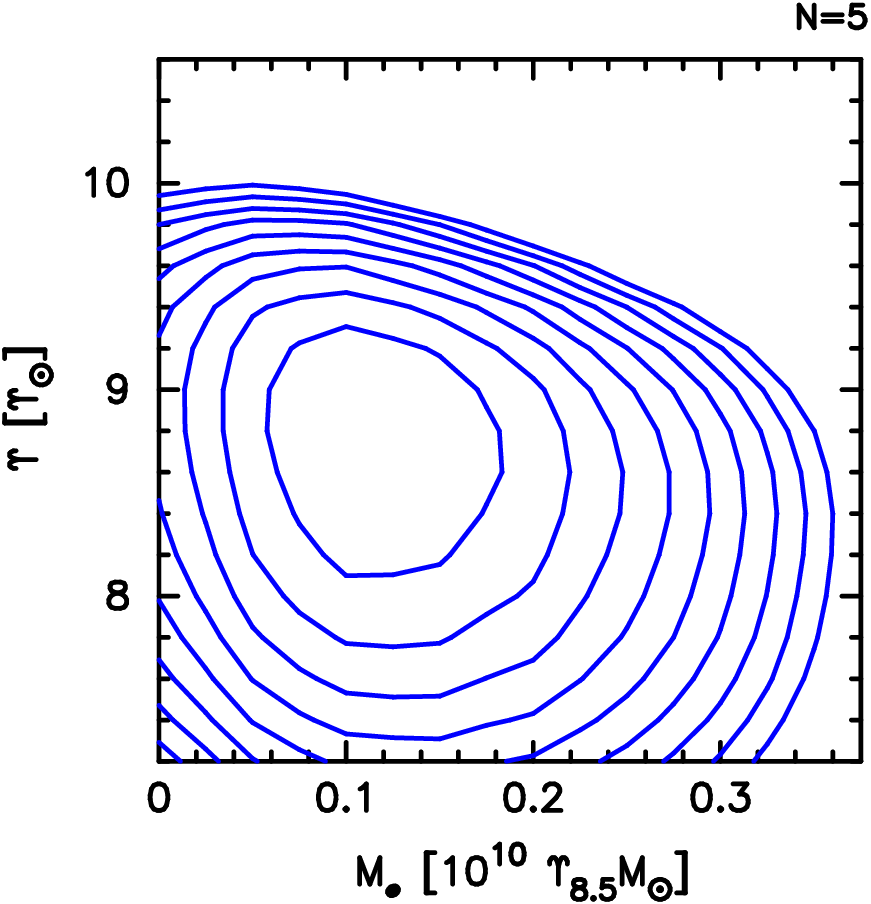}
  \hfil
  \includegraphics[width=0.4\hsize,angle=270]{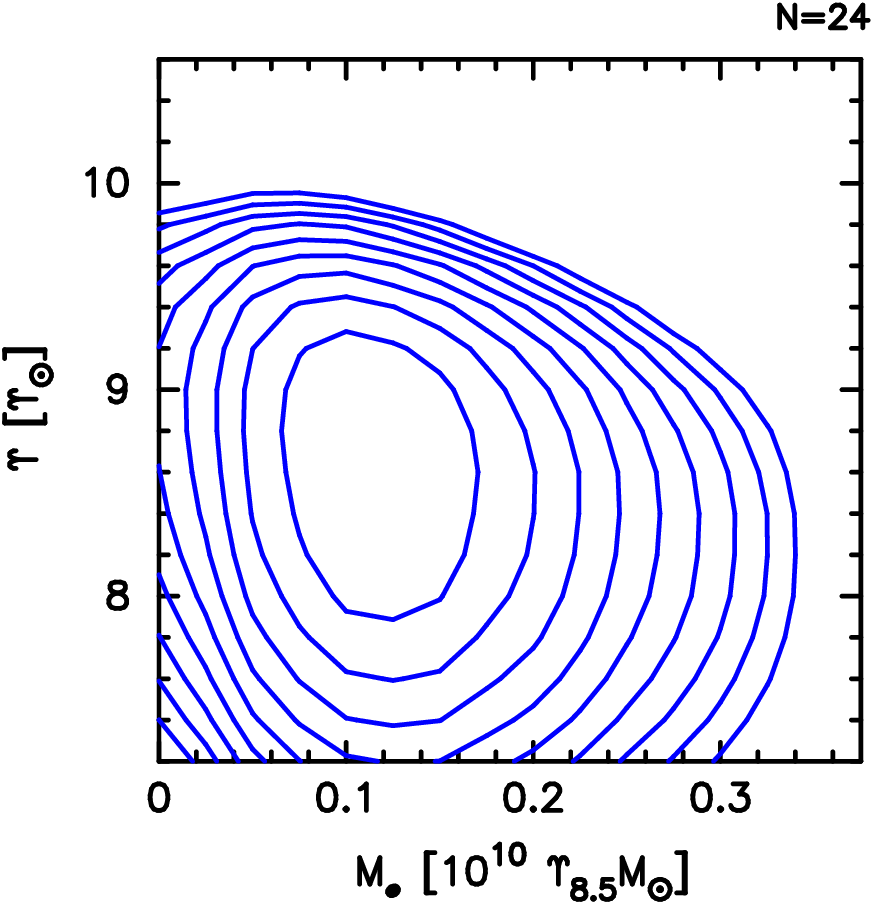}}
\caption{As for fig.~\ref{fig:chisqpotgh}, but fitting to eigenVPs
  (\S\ref{sec:modelfithist}) instead of Gauss--Hermite coefficients.
  The first three panels show the results of fitting only to the first
  3, 4 and 5 eigenVPs of each spectrum.  The last shows the results of
  fitting to the full spectrum by using all 24 eigenVPs. }
\label{fig:chisqpothist}
\end{figure*}

We have calculated the projection coefficients $L_{ij}^{(k)}(\psi)$
and $I_i^{(k)}(\psi)$ for $n_\E\times n_J=200\times40$ DF components
(eq.~\ref{eq:DFdoublesum}) in a range of potentials~$\psi$.  Our main
results below are obtained using coefficients calculated for BH masses
$M_\bullet/10^9M_\odot=0,0.25,\ldots,3.75$ with a single mass-to-light
ratio, $\Upsilon_0=8.5\Upsilon_\odot$.  Since the projection
coefficients scale straightforwardly with mass, we can use the method
of~\S\ref{sec:modelfit} to fit models with other values of $\Upsilon$
provided we remember to scale $M_\bullet$ by $\Upsilon/\Upsilon_0$ and
the bins~$v_j$ of the velocity histograms~(\ref{eq:vpcoeff}) by
$\sqrt{\Upsilon/\Upsilon_0}$.

On figure~\ref{fig:chisqpotgh} we plot the result of
using~(\ref{eq:chisqmodall}) to fit the Gauss--Hermite coefficients
$(h_0,h_2,\ldots,h_N)$ for the cases $N=4$ and $N=6$.  In each case we
have extracted VPs from spectra assuming that $h_i=0$ for $i>N$.
Marginalizing~$\Upsilon$, the former yields a moderately firm BH mass
$M_\bullet\simeq2.3^{+1.1}_{-0.9}\times10^9\,M_\odot$.  Including
$h_6$ in the fit, however, results only in the upper bound,
$M_\bullet<1.6\times10^9\,M_\odot$, statistically consistent with the
$N=4$ result.  To test whether this could be a result of the best-fit
template fraction~$f$, continuum~$c_i$ or normalisation~$k$
(Sec.~\ref{exkin}) changing as we change~$N$, we have repeated our
$N=6$ model fit omitting the coefficient $h'_j$ (eq.~\ref{eq:sGH})
with the largest eigenerror~$\Delta_j$.  The resulting~$\chi^2_{\rm
  m}(M_\bullet,\Upsilon)$ contours are broadly similar to the $N=4$
case, suggesting that the shift in the location of the minimum
in~$\chi^2_{\rm m}$ is a genuine effect caused by the extra
information contained in the $h_6$ parameter, and not by systematic
changes in the fitted continuum or normalisation.

These results are perhaps unsurprising when one recalls that our
results in Sec. \ref{sec:results} indicate that NGC~1399 has very strongly
non-Gaussian central VPs.  Although it would be a mildly interesting
exercise to test the effects of fitting to $h_8$ and even higher-order
terms, we now simply drop the Gauss--Hermite parametrisation and turn
to fitting eigenVPs.  Fitting models to the full set of eigenVPs
yields the result plotted in the bottom-right of
fig.~\ref{fig:chisqpothist}.  The best-fit BH mass,
$1.2^{+0.5}_{-0.6}\times10^9\,M_\odot$, is consistent with the results from
the Gauss--Hermite fits, but has smaller error bars.  The other panels
on the figure show the effect of rearranging the eigenVPs in order of
increasing eigenerror and fitting only the first $N$ for each
spectrum.  The case $N=3$ yields results that look qualitatively
similar to our sixth-order Gauss--Hermite fit, while adding one more
eigenVP introduces a useful lower-bound on the BH mass.  The results
for $N=5$ are very similar to fitting the full $N=24$: in fact, it is
impossible to distinguish between $N=6$ and $N=24$ by eye.

\begin{figure}
\includegraphics[width=0.6\hsize,angle=270]{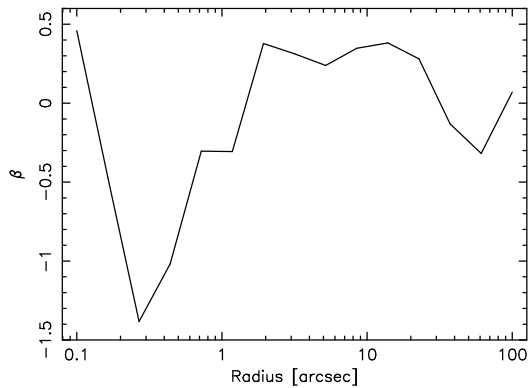}
\caption{Anisotropy parameter $\beta\equiv
  1-\sigma_\theta^2/\sigma_r^2$ of a typical model
  ($M_\bullet=10^9M_\odot$, $\Upsilon=8.5\Upsilon_\odot$), averaged
  over five shells per decade in radius.}
\label{fig:aniso}
\end{figure}
Since the the $N=24$ model is essentially a direct fit to the galaxy
spectrum we adopt its best-fit
$M_\bullet=1.2^{+0.5}_{-0.6}\times10^9\,M_\odot$ as our best estimate of the
BH mass in NGC~1399.  Figure~\ref{fig:aniso} plots the anisotropy
parameter of one of the best-fit models in this range.  Our models
have moderate radial anisotropy ($\beta\approx0.3$) between 2 arcsec
and 30 arcsec, similar to the results found by \citet{saglia00} in
their models of NGC~1399.  At larger radii the orbit distribution in
our models becomes tangentially biased, which is what one expects from
fitting a constant mass-to-light ratio model to a galaxy with a
massive dark halo.  Much more surprisingly, however, our models become
extremely tangentially biased in the innermost arcsec, which is where we
find the interesting kinematic and photometric features.

\subsection{Tests}
\label{sec:modeltests}
Our best-fit BH mass remains unchanged if we change the number of DF
components we use in the models: models using $100\times20$ components
instead of $200\times40$ yield the same result.  We find no evidence
for the flat-bottomed $\chi^2_{\rm m}$ profiles claimed by
\citet{vme04} (albeit for the axisymmetric case), even though our
models sample $(\E,J^2)$ phase space a factor $\sim20$ more densely
than theirs.  We have tried varying the extent and widths of the
velocity bins we use when we fit VPs, but the BH mass is unchanged
whether we take 24 fine 25\kms-wide bins extending to 1200\kms or 20
coarse 100\kms-wide bins extending to 2000\kms.

The assumed width of the velocity bins does affect the models in a
slightly more subtle way, however.  When we rescale our base
$\Upsilon_0=8.5\Upsilon_\odot$ models to a new mass-to-light
ratio~$\Upsilon$, we scale their velocity bins by an
amount~$\sqrt{\Upsilon/\Upsilon_0}$ and re-extract VPs from spectra
using the the new bin widths.  Therefore, values of $\chi^2_{\rm s}(\b
L)$ from~(\ref{eq:chisqhistvp}) for different~$\Upsilon$ are not
strictly directly comparable.  To test the effects of this we have
calculated coarser $100\times20$-component models using 24 bins of
fixed width 50\kms on a grid of potentials with
$\Upsilon/\Upsilon_\odot$ running from 7.2 to 10.6 in 0.2 steps and
$M_\bullet/10^9M_\odot$ running from 0 to 3.75 in 0.25 steps.  The
only notable difference between the resulting $\chi^2_{\rm
  m}(M_\bullet,\Upsilon)$ and the plots in fig.~\ref{fig:chisqpothist}
is that the range of acceptable $\Upsilon$ increases to
$(9\pm1)\Upsilon_\odot$.  The BH mass is unaffected.

\subsection{Caveats on BH mass}
\label{sec:modelcaveats}
Despite these reassurances, there nevertheless are some shortcomings
of the models and data upon which this BH mass is based:
\begin{enumerate}
\item We follow the usual ``extended-Schwarzschild'' procedure from
  which the majority of existing stellar-dynamical BH masses have been obtained
  and consider only the very best orbit distribution $f_k$ for each
  potential.  This best-fit distribution is typically very spiky, with
  only $\sim140$ non-zero~$f_k$ out of $200\times40$!  One way around this
  would be to apply some kind of regularization to the~$f_k$
  \citep{jens05}, but the biases introduced by this procedure are not
  well understood \citep{vme04}.
\item Because of the huge freedom in fitting the~$f_k$, the best-fit
  model should have a very low $\chi^2_{\rm m}$.  For example, in
  Monte Carlo experiments with synthetic datasets of toy galaxy models
  we typically find $\chi^2_{\rm m}\sim N_{\rm data}/3$, where $N_{\rm
    data}$ is the number of data points fit in the models.  For our
  real NGC~1399 data, however, our best-fit $\chi^2_{\rm GH}$ and
  $\chi^2_{\rm H}$ are usually at least as big as the number of
  parameters we use to describe our kinematics.  This is probably due
  to our neglect of the systematic errors
  (Sec. \ref{telcor} and \ref{sec:fd}).  In contrast, the fit to the full surface brightness
  profile and to the outer dispersion profile is astonishingly good,
  with $\chi^2_I+\chi^2_G\sim3$, making our total $\chi^2_{\rm m}$
  relatively low.
\item Finally, our models assume that galaxy is spherical and
  non-rotating, despite the clear evidence to the contrary.
  Nevertheless, the models do seem to require a strong bias towards
  circular orbits in the central 0.5 arcsec.
\end{enumerate}

\section{Conclusions}
\label{sec:conc}
Using NAOS-CONICA at the VLT, we have successfully measured the central 
kinematics within the SoI of NGC 1399 ($r \sim$ 0\farcs34) with a resolution 
(FWHM) of $\sim$0\farcs15 (14pc, Sec \ref{sec:AO}) using adaptive optics correction 
on a bright reference star 17\farcs5 away.

Alone, the kinematics extracted from the CaI feature at 2.26\mum\ and the 
CO bands after 2.3\mum\ establish the presence of velocity gradient within 
a radius of $\sim$0\farcs5 (48pc), suggestive of a 
kinematically decoupled core.

Ks band imaging reveals offset and elongated isophotes within a radius of 
0\farcs2 (19pc) that are not visible in H-band HST images. Such inner 
structure is reminiscent of that seen in other core ellipticals. The 
non-parametric VPs corresponding to this region also show an unusual 
velocity structure that may be consistent with the presence of an 
eccentric disk around the BH, akin to that of M31.

We have demonstrated that errors in the Gauss--Hermite coefficients~$h_j$ 
extracted from real galaxy spectra are {\it not} independent, and have 
shown a simple way of taking the covariances among the~$h_j$ into account 
when fitting dynamical models.  The VPs near the nucleus of NGC~1399 are 
strongly non-Gaussian, however, and are not well described by a low-order 
Gauss--Hermite expansion.  We show that the ``eigenVPs'' obtained by 
diagonalizing~(\ref{eq:chisqspec}) are a more useful way of describing 
VPs, at least for numerical purposes.

Subject to the caveats of \S6.6, our best estimate for the mass of the
BH in NGC~1399 is $1.2^{+0.5}_{-0.6}\times10^9M_\odot$, obtained by
fitting spherical dynamical models directly to our observed
spectra. The models are based on the usual extension of
Schwarzschild's method to the problem of potential estimation. Taken
at face value, they place this galaxy on the $M_\bullet$-$\sigma$
plane mid-way between the predictions of T02 and FF05, being
consistent with both. The best-fit model also becomes extremely
tangentially anisotropic in the innermost 0\farcs5.

We have demonstrated that AO observations are a viable alternative to
HST to when measuring black-hole masses and can break the
mass-anisotropy degeneracy even in the most massive, non-rotating
elliptical galaxies. The difficulty in interpreting these long-slit
data emphasises the need for high SNR AO assisted integral-field
observations to further understand the kinematic and photometric
features discovered here.

\section*{Acknowledgements}
We thank the referee, Laura Ferrarese for her constructive comments. This research is based on observations collected at the European Southern Observatory, Chile (ESO Program 072.B-0763). We acknowledge use of the SIMBAD Astronomical Database and the HyperLeda database. Authors are funded by the Particle Physics and Astronomy Research Council (PPARC) and the Royal Society.

\bsp

\label{lastpage}


\begin{thebibliography}{99}
\bibitem[\protect\citeauthoryear{Aller \& Richstone}{2002}]{AR2002} Aller, M.~C., \& Richstone, D.\ 2002, \aj, 124, 3035 
\bibitem[\protect\citeauthoryear{Baird}{1981}]{b1} Baird S.R., 1981,ApJ, 245, 208
\bibitem[\protect\citeauthoryear{Bender}{1990}]{Bender90} Bender, R.\ 1990, \aap, 229, 441 
\bibitem[\protect\citeauthoryear{Bicknell, Bruce, Carter \& Killeen}{1989}]{bick89} Bicknell, G.~V., Bruce, T.~E.~G., Carter, D., \& Killeen, N.~E.~B. 1989, ApJ, 336, 639 
\bibitem[\protect\citeauthoryear{Binney \& Mamon}{1982}]{binney&mamon82} Binney, J.~\& Mamon, G.~A.\ 1982, MNRAS, 200, 361 
\bibitem[\protect\citeauthoryear{Burstein et al.}{1987}]{bustein87} Burstein, D., Davies, R.~L., Dressler, A., Faber, S.~M., Stone, R.~P.~S., Lynden-Bell, D., Terlevich, R.~J., \& Wegner, G.\ 1987, \apjs, 64, 601 

\bibitem[\protect\citeauthoryear{Cappellari \& Emsellem}{2004}]{cap&em04} Cappellari, M., \& Emsellem, E.\ 2004, \pasp, 116, 138 
\bibitem[\protect\citeauthoryear{Caon et al.}{1994}]{Caon94} Caon, N., Capaccioli, M., \& D'Onofrio, M.\ 1994, \aaps, 106, 199 
\bibitem[\protect\citeauthoryear{Cretton et al.}{1999}]{cretton99} Cretton N., de Zeeuw P.~T., van der Marel 
R.~P., Rix H.-W., 1999, ApJS, 124, 383 
\bibitem[\protect\citeauthoryear{Efstathiou et al.}{1980}]{EfEC1980} Efstathiou, G., Ellis, R.~S., \& Carter, D.\ 1980, \mnras, 193, 931 
\bibitem[\protect\citeauthoryear{Emsellem et al.}{2004}]{sauronIII} Emsellem, E., et al.\ 2004, \mnras, 352, 721
\bibitem[\protect\citeauthoryear{Faber et al.}{1997}]{faber97} Faber, S.~M., et al. 1997, AJ, 114, 1771 
\bibitem[\protect\citeauthoryear{Ferrarese \& Merrit}{2000}]{f&m2000} Ferrarese, L.~\& Merritt, D. 2000, ApJL, 539, L9 
\bibitem[\protect\citeauthoryear{Ferrarese \& Ford}{2005}]{FF05} Ferrarese, L., \& Ford, H.\ 2005, Space Science Reviews, 116, 523 
\bibitem[\protect\citeauthoryear{Forbes et al.}{1994}]{Forbes94} Forbes, D.~A., Franx, M., \& Illingworth, G.~D.\ 1994, \apjl, 428, L49 
\bibitem[\protect\citeauthoryear{Franx, Illingworth \& Heckman}{1989}]{f&i&h89} Franx, M., Illingworth, G., \& Heckman, T. 1989, ApJ, 344, 613 
\bibitem[\protect\citeauthoryear{Gerhard}{1993}]{Gerhard93} Gerhard, O.~E.\ 1993, \mnras, 265, 213 
\bibitem[\protect\citeauthoryear{Gebhardt}{2000}]{geb2000} Gebhardt, K., et al. 2000, ApJL, 539, L13 
\bibitem[\protect\citeauthoryear{Gebhardt et al.}{2003}]{geb03} Gebhardt K., et al., 2003, ApJ, 583, 92 

\bibitem[\protect\citeauthoryear{Graham et al.}{1998}]{graham98} Graham, A.~W., Colless, M.~M., Busarello, G., Zaggia, S., \& Longo, G.\ 1998, \aaps, 133, 325 
\bibitem[\protect\citeauthoryear{Haehnelt \& Kauffmann}{2000}]{H&K2000} Haehnelt, M.~G., 
\& Kauffmann, G.\ 2000, \mnras, 318, L35
\bibitem[\protect\citeauthoryear{H{\" a}ring \& Rix}{2004}]{HR2004} H{\" a}ring, N., \& Rix, H.-W.\ 2004, \apjl, 604, L89 

\bibitem[\protect\citeauthoryear{Killeen \& Bicknell}{1988}]{kill&bick88} Killeen, N.~E.~B., 
\& Bicknell, G.~V.\ 1988, \apj, 325, 165 
\bibitem[\protect\citeauthoryear{Killeen et al.}{1988}]{kill_et_al88} Killeen, N.~E.~B., 
Bicknell, G.~V., \& Ekers, G.~V.\ 1988, \apj, 325, 180 
\bibitem[\protect\citeauthoryear{Kleinmann \& Hall}{1986}]{k&h86} Kleinmann, S.~G., \& Hall, D.~N.~B.\ 1986, \apjs, 62, 501 
\bibitem[\protect\citeauthoryear{Kormendy \& Richstone}{1995}]{kor&rich95} Kormendy, J.~\& Richstone, D. 1995, ARA\&A, 33, 581 
\bibitem[\protect\citeauthoryear{Kuntschner}{2000}]{K2000} Kuntschner, H. 2000, MNRAS, 315, 184 
\bibitem[\protect\citeauthoryear{Landsman}{1993}]{landsman93} Landsman, W.~B.\ 1993, ASP Conf.~Ser.~ 52: Astronomical Data Analysis Software and Systems II, 52, 246 
\bibitem[\protect\citeauthoryear{Lawson \& Hanson}{1974}]{LH1974} Lawson, C. L., \& Hanson, R. J. 1974, Solving Least Squares Problems (Englewood Cliffs, New Jersey: Prentice-Hall)
\bibitem[\protect\citeauthoryear{Lauer et al.}{2005}]{Lauer05} Lauer, T.~R., et al.\ 2005, \aj, 129, 2138 
\bibitem[\protect\citeauthoryear{Lauer et al.}{1995}]{Lauer95} Lauer, T.~R., et al.\ 1995, \aj, 110, 2622 

\bibitem[\protect\citeauthoryear{Leitherer et al.}{1999}]{star99} Leitherer, C., et al. 1999, ApJS, 123, 3 
\bibitem[\protect\citeauthoryear{Lenzen et al.}{1998}]{conica} Lenzen, R., Hofmann, R., Bizenberger, P., \& Tusche, A.\ 1998, \procspie, 3354, 606 
\bibitem[\protect\citeauthoryear{Longo et al.}{1994}]{lon94} Longo, G., Zaggia, S., Busarello, G., \& Richter, G.\ 1994, VizieR Online Data Catalog, 410, 50433 
\bibitem[\protect\citeauthoryear{Madore et al.}{1999}]{distance} Madore, B.~F., et al.\ 1999, \apj, 515, 29


\bibitem[\protect\citeauthoryear{Magorrian \& Binney}{1994}]{mag_bin94} Magorrian, J.~\& Binney, J. 1994, MNRAS, 271, 949
\bibitem[\protect\citeauthoryear{Magorrian et al.}{1998}]{mag98} Magorrian, J., et al. 1998, AJ, 115, 2285 
\bibitem[\protect\citeauthoryear{Magorrian et al.}{2005}]{mag05} Magorrian J., et al., in preparation
\bibitem[\protect\citeauthoryear{Marconi \& Hunt}{2003}]{m&h2003} Marconi, A., \& Hunt, L.~K.\ 2003, \apjl, 589, L21
\bibitem[\protect\citeauthoryear{van der Marel \& Franx}{1993}]{vdM&Franx93} van der Marel, R.~P.~\& Franx, M. 1993, ApJ, 407, 525 
\bibitem[\protect\citeauthoryear{van der Marel et al.}{1994}]{vdMetal94} van der Marel, R.~P., Rix, H.~W., Carter, D., Franx, M., White, S.~D.~M., \& de Zeeuw, T.\ 1994, \mnras, 268, 521 
\bibitem[\protect\citeauthoryear{van der Marel}{1994}]{vdM94} van der Marel, R.~P.\ 1994, \mnras, 270, 271
\bibitem[\protect\citeauthoryear{Merrit \& Ferrarese}{2001}]{m&f2001} Merritt, D.~\& Ferrarese, L. 2001, MNRAS, 320, L30 
\bibitem[\protect\citeauthoryear{Oliva \& Origlia}{1992}]{airglow} Oliva, E., \& Origlia, L.\ 1992, \aap, 254, 466 

\bibitem[\protect\citeauthoryear{Origlia, Moorwood and Oliva}{1993}]{omo93} Origlia, L., Moorwood, A.~F.~M., \& Oliva, E. 1993, A\&A, 280, 536 
\bibitem[\protect\citeauthoryear{Oliva et al.}{1995}]{oliva95} Oliva, E., Origlia, L., Kotilainen, J.~K., \& Moorwood, A.~F.~M.\ 1995, \aap, 301, 55 
 
\bibitem[\protect\citeauthoryear{D'Onofrio et al.}{1995}]{don95} D'Onofrio, M., Zaggia, S.~R., Longo, G., Caon, N., \& Capaccioli, M.\ 1995, \aap, 296, 319 

\bibitem[\protect\citeauthoryear{Pinkney et al.}{2003}]{pink2003} Pinkney, J., et al.\ 2003, \apj, 596, 903 

\bibitem[\protect\citeauthoryear{Puxley, Doyon and Ward}{1997}]{pdw97} Puxley, P.~J., Doyon, R., \& Ward, M.~J. 1997, ApJ, 476, 120 
\bibitem[\protect\citeauthoryear{Richstone \& Sargent}{1972}]{r&s72} Richstone, D.~\& Sargent, W.~L.~W. 1972, ApJ, 176, 91 
\bibitem[\protect\citeauthoryear{De Rijcke \& Dejonghe}{1998}]{DD98} De Rijcke S., Dejonghe H., 1998, MNRAS, 298, 677 

\bibitem[\protect\citeauthoryear{Rix \& White}{1992}]{rix&white92} Rix, H.~\& White, S.~D.~M. 1992, MNRAS, 254, 389 
\bibitem[\protect\citeauthoryear{Rix et al.}{1997}]{rix97} Rix H.-W., de Zeeuw P.~T., Cretton N., van der Marel R.~P., Carollo C.~M., 1997, ApJ, 488, 702 
\bibitem[\protect\citeauthoryear{Rousset et al.}{1998}]{naos} Rousset, G., et al.\ 1998, \procspie, 3353, 508 
\bibitem[\protect\citeauthoryear{Saglia et al.}{2000}]{saglia00} Saglia, R.~P., Kronawitter, A., Gerhard, O., \& Bender, R.\ 2000, ApJ, 119, 153 
\bibitem[\protect\citeauthoryear{Saha \& Williams}{1994}]{saha&will94} Saha, P., \& Williams, T.~B.\ 1994, \aj, 107, 1295 
\bibitem[\protect\citeauthoryear{Sargent et al.}{1978}]{sargent78} Sargent, W.~L.~W., Young, P.~J., Lynds, C.~R., Boksenberg, A., Shortridge, K., \& Hartwick, F.~D.~A. 1978, ApJ, 221, 731 
\bibitem[Sch{\" o}del et al.(2003)]{Schodel2003} Sch{\" o}del, R., Ott, T., Genzel, R., Eckart, A., Mouawad, N., \& Alexander, T.\ 2003, \apj, 596, 1015 
\bibitem[\protect\citeauthoryear{Schombert}{1986}]{schom86} Schombert, J.~M.\ 1986, 
\apjs, 60, 603 
\bibitem[\protect\citeauthoryear{Simkin}{1974}]{simkin74} Simkin, S.~M. 1974, A\&A, 31, 129 
\bibitem[\protect\citeauthoryear{Smith et al.}{2000}]{smith2000} Smith, R.~J., Lucey, J.~R., Hudson, M.~J., Schlegel, D.~J., \& Davies, R.~L. 2000, MNRAS, 313, 469 
\bibitem[\protect\citeauthoryear{Thatte, Tecza and Genzel}{2000}]{ttg2000} Thatte, N., Tecza, M., \& Genzel, R. 2000, A\&A, 364, L47 
\bibitem[\protect\citeauthoryear{Thomas et al.}{2005}]{jens05} Thomas J., Saglia R.~P., Bender R., Thomas D., Gebhardt K., Magorrian J., Corsini E.~M., Wegner G., 2005, MNRAS, 360, 1355 
\bibitem[\protect\citeauthoryear{Tonry \& Davis}{1979}]{tonry&davis79} Tonry, J.~\& Davis, M. 1979, AJ, 84, 1511 
\bibitem[\protect\citeauthoryear{Tonry et al.}{2001}]{tonry2001} Tonry, J.~L., Dressler, A., Blakeslee, J.~P., Ajhar, E.~A., Fletcher, A.~B., Luppino, G.~A., Metzger, M.~R., \& Moore, C.~B.\ 2001, \apj, 546, 681 
\bibitem[\protect\citeauthoryear{Tremaine}{1995}]{tremaine95} Tremaine, S.\ 1995, \aj, 110, 628 
\bibitem[\protect\citeauthoryear{Tremaine et al.}{2002}]{T02} Tremaine, S., et al. 2002, ApJ, 574, 740 
\bibitem[\protect\citeauthoryear{Turner, Ho and Beck}{1987}]{thb87} Turner, J.~L., Ho, P.~T.~P., \& Beck, S. 1987, ApJ, 313, 644 
\bibitem[\protect\citeauthoryear{Valluri, Merritt, \& Emsellem}{2004}]{vme04} Valluri M., Merritt D., Emsellem E., 2004, ApJ, 602, 66 
\bibitem[\protect\citeauthoryear{de Vaucouleurs et al.}{1991}]{vau91} de Vaucouleurs, G., de Vaucouleurs, A., Corwin, H.~G., Buta, R.~J., Paturel, G., \& Fouque, P. 1991, Volume 1-3, XII, 2069 pp.~7 figs..~ Springer-Verlag Berlin Heidelberg New York,  
 
\bibitem[\protect\citeauthoryear{Vega et al.}{2001}]{vega2001} Vega Beltr{\' a}n, J.~C., Pizzella, A., Corsini, E.~M., Funes, J.~G., Zeilinger, W.~W., Beckman, J.~E., \& Bertola, F. 2001, A\&A, 374, 394 
\bibitem[\protect\citeauthoryear{Yu \& Tremaine}{2002}]{YT2002} Yu, Q., \& Tremaine, S.\ 2002, \mnras, 335, 965 \bibitem[\protect\citeauthoryear{de Zeeuw}{2003}]{zeeuw2003} de Zeeuw, T., astro-ph/0303469
\bibitem[\protect\citeauthoryear{Zhao et al.}{2002}]{Zhao02} Zhao, H., Haehnelt, M.~G., \& Rees, M.~J.\ 2002, New Astronomy, 7, 385 


\end{thebibliography}
\end{document}